\documentclass[sigconf]{acmart}

\usepackage{booktabs} % For formal tables

\setcopyright{rightsretained}
\begin{document}
\title{Toward the Understanding of Deep Text Matching Models for Information Retrieval}

\author{Lijuan Chen$^1$, Yanyan Lan$^{2*}$, Liang Pang$^3$, Jiafeng Guo$^4$, Xueqi Cheng$^4$}

\email{lanyanyan@tsinghua.edu.cn, chenlijuan@sogou-inc.com, {pangliang, guojiafeng, cxq}@ict.ac.cn}

\affiliation{%
   \institution{
   $^1$Sogou Inc., Beijing, China \\
   $^2$Institute for AI Industry Research, Tsinghua University, Beijing, China \\
   $^3$Data Intelligence System Research Center, \\
   Institute of Computing Technology, Chinese Academy of Sciences, Beijing, China \\
   $^4$CAS Key Lab of Network Data Science and Technology, \\
   Institute of Computing Technology, Chinese Academy of Sciences, Beijing, China \\
   }
}

\begin{abstract}
Semantic text matching is a critical problem in information retrieval. Recently, deep learning techniques have been widely used in this area and obtained significant performance improvements. However, most models are black boxes and it is hard to understand what happened in the matching process, due to the poor interpretability of deep learning. This paper aims at tackling this problem. The key idea is to test whether existing deep text matching methods satisfy some fundamental heuristics in information retrieval. Specifically, four heuristics are used in our study, i.e., term frequency constraint, term discrimination constraint, length normalization constraints, and TF-length constraint. Since deep matching models usually contain many parameters, it is difficult to conduct a theoretical study for these complicated functions. In this paper, We propose an empirical testing method. Specifically, We first construct some queries and documents to make them satisfy the assumption in a constraint, and then test to which extend a deep text matching model trained on the original dataset satisfy the corresponding constraint. Besides, a famous attribution based interpretation method, namely integrated gradient, is adopted to conduct detailed analysis and guide for feasible improvement. Experimental results on LETOR 4.0 and MS Marco show that all the investigated deep text matching methods, both representation and interaction based methods, satisfy the above constraints with high probabilities in statistics. We further extend these constraints to the semantic settings, which are shown to be better satisfied for all the deep text matching models. These empirical findings give clear understandings on why deep text matching models usually perform well in information retrieval. We believe the proposed evaluation methodology will be useful for testing future deep text matching models.
\end{abstract}


%\begin{CCSXML}
%<ccs2012>
%<concept>
%<concept_id>10010147.10010178.10010179</concept_id>
%<concept_desc>Computing methodologies~Natural language processing</concept_desc>
%<concept_significance>300</concept_significance>
%</concept>
%</ccs2012>
%\end{CCSXML}
%\ccsdesc[300]{Computing methodologies~Natural language processing}

%\keywords{Interpretation, Semantic Matching, Integrated Gradients}

\maketitle
\section{Introduction}
Matching two sentences is a core problem in many information retrieval (IR) applications, such as web search \cite{kelly2009foundations}, question answering \cite{berger2000bridging} and paraphrasing identification\cite{dolan2004unsupervised}. Take web search as an example, given a query and a document, the matching score is usually used to determine the relevance between them.
 
Recently, deep neural networks have been widely applied in this area and achieved great progresses. These deep text matching models are usually divided into two categories, i.e.~representation and interaction based methods. The representation based methods use deep neural network to obtain the representations of each object, then conduct the interaction information to output a matching score. Representative models includes DSSM \cite{huang2013learning}, CDSSM \citep{shen2014latent} and ARC-I \citep{hu2014convolutional}, etc. While the interaction based methods directly apply deep neural networks on the interaction matrix of the two objects to output the matching score. DRMM \citep{Guo2016A}, MatchPyramid \citep{pang2016text}, KNRM\citep{xiong2017end}, DeepMatch\citep{revaud2016deepmatching} and ARC-II\citep{hu2014convolutional} have been recognized as some typical models of this kind. More recently, the popular pre-training technique, e.g.~BERT~\citep{devlin2018bert} has also been applied to deep text matching models and gain increasing attention.

Though deep text matching models have shown good performance in information retrieval, it is unclear what happened in matching process. Features are not explicit any more in these deep text matching models, as compared with the traditional learning-to-rank methods. Deep learning models are black boxes themselves. Therefore, it is very hard to understand why deep text matching models perform well, and what kind of knowledge/principles do they learn or capture in the matching process. This is exactly the motivation of this paper. We should note that this problem is very challenging. Firstly, interpretation itself is a difficult problem for in the field of deep learning, though the direction has obtained significant attentions and several different interpretation methods have been proposed, such as feature visualizations \citep{olah2017feature}, attribution methods \citep{ancona2017unified} , and sample importance methods \citep{koh2017understanding}. Secondly, the formal definition of interpretability is not clear, and may differ for various applications.

Looking back at IR history, some IR heuristics, i.e., several basic desirable constraints, have been  proposed in \cite{fang2004formal, fang2005exploration}. The performance of a retrieval formula is tightly related to how well it satisfies these constraints. Inspired by this finding, we propose to conduct the understanding of existing deep text matching methods from the perspective of IR heuristics, including term frequency constraints (TFCs) \cite{Salton1987Term}, term discrimination constraint (TDC), length normalization constraints (LNCs) \cite{Zobel1998Exploring}, and TF-length constraint (TF-LNC). We noticed that \citep{rennings2019axiomatic} has conducted a similar empirical study. However, they focus on diagnosing the deep model whether they can be improved by adding some data which satisfy the assumption of the constraints. Therefore, there are two problems in this approach: 1) it fails to detect whether these constraints are truly satisfied by a deep text matching model; 2) comparisons between different deep learning models are not allowed.

To address these limitations, this paper focus on study whether these deep text matching models satisfy the existing IR constraint. Since the deep text matching models are usually very complicated and contain many parameters, it is not feasible any more to directly conduct mathematical derivations to achieve the conclusion. So we propose to test the trained models on constructed test data. Firstly, we train a deep text matching model on training data. Then we construct queries and documents which satisfy the assumption of a constraint to form a test data. Finally, the trained model is applied on test data, and the proportion of data that satisfy the constraint can be obtained. This value reflect to which extent the deep text matching model satisfy this constraint. Furthermore, the interpretation method Integrated Gradient (IG) \citep{sundararajan2017axiomatic}, which has been proven to be stable and reliable in many different applications, is used in our experiments to conduct detailed analysis and improvements.

We experiment  on two widely used datasets in IR, i.e.~LETOR 4.0 and MS Marco. Three kinds of deep text matching models are tested, including representation based methods such as ARC-I, interaction based methods such as MatchPyramid, KNRM and BERT, and the hybrid models such as RI-Match and DUET. The results show that these deep text matching models satisfy the four concerned constraints with high probabilities in statistics, which explain why deep text matching models usually perform well on many IR tasks. Furthermore, we extend the above constraints to the semantic versions, by incorporating the word embeddings into the definitions. Experiments show that the deep text matching models satisfy semantic constraints with higher probabilities, which explains the mechanism of how these models capture the semantic matching relations between queries and documents in the scenario of IR.

Our main contributions include: 1) the proposal of a method to test whether a deep text matching model satisfies the existing IR heuristics, which can be used for existing and future deep learning models; 2) the extensive empirical studies on LETOR 4.0 and MS Marco, including both representation and interaction based models; 3) the extension of existing IR constraints to the semantic versions, which provide some foundations for potential investigations of modern deep learning based retrieval models.

\section{Backgrounds}
In this section, we introduce backgrounds that including existing deep matching models for IR, and the interpretation method used in this paper, i.e., integrated gradient (IG).
\subsection{Deep Text Matching Models}
Recently, deep text matching technique has been widely applied in IR, and existing models can be mainly divided into two categories, i.e representation based methods and interaction based methods.

Representation based methods focus on representing query and document to two vectors by using different deep neural networks, such as CNNs \cite{kalchbrenner2014convolutional,denil2014modelling} and RNNs \citep{li2015hierarchical,touretzky1996advances}. Then matching score is computed by similarity function or multiple layer perceptron(MLP). Typical representation based  models include DSSM, CDSSM, ARC-I, LSTM-RNN\citep{palangi2016deep}. DSSM adopts a feed forward neural network with letter  trigram representation as the input. CDSSM and ARC-I both represent the input by CNN. For CDSSM, the input format is  letter trigram representation, while ARC-I is a CNN with word embeddings as the input. LSTM-RNN  utilize RNN embeds document into a semantic vector. In general, this approach is straightforward and capture the high level semantic meanings of each sentence.

Though representation based models are easy to understand and implement, they usually lose rich detailed interaction features. Interaction based models have been proposed to overcome shortcoming. Therefore, a matching matrix is firstly used to capture the word level query-document interaction features. Then different deep neural networks are utilized to further capture the high level matching features. At last, similar to representation based models, the matching score is produced by a simple similarity function or a MLP. Typical interaction based text matching models include ARC-II
\citep{hu2014convolutional} , MatchPyramid\citep{pang2016text}, Match-SRNN\cite{wan2016match}, KNRM\citep{xiong2017end} and BiMPM\cite{wang2017bilateral}. In ARC-II and MatchPyramid, interaction information is calculated by a mapping function to map query/document to a sequence of word representations, then ARC-II adopts  1-D CNN structure to scan each patch of words from query and document, while MatchPyramid adopts CNN to obtain it. Match-SRNN utlize  tensor operation to incorporate high dimensional word level interactions, then  2D-GRU structure used to process the information. In KNRM, the translation layer calculates the word-word similarities to form the translation matrix, the kernel pooling process above matrix. BiMPM utilize multi-perspective matching operation including the attentive  matching to capture the interaction information. BERT obtain the interaction information by the Transformert structure.

Both interaction information and text representations are needed to determine the matching score. To further improve performance of deep text matching models, DUET\citep{mitra2017learning} and  RI-Match \citep{chen2018ri} are proposed to combine the merits of both deep matching approaches to improve the performance of text matching.

Although the existing deep text matching models have achieved great success in many IR tasks, models are still black boxes for us. The understandings of these models are critical because they can not only help explain how these model work, but also give some insights on how to design better models. However, rare studies have been conducted in this area. The only work on this topic is \citep{rennings2019axiomatic}, which is very similar to us because they also conduct an empirical study for deep text matching models on IR heuristics. However, it should be noted that our approach are quite different from them. They mainly diagnose a deep model by adding data satisfying the constraint. If a model achieves performance improvements on added data, it is recognized as a good model. However, this approach cannot truly determine whether a model satisfy the IR constraints. Furthermore, using the performance improvement for a single model on different data fail to achieve an comparison between different models. Our work addresses these two limitations. In addition, we adopt an interpretation algorithm to conduct a detailed data analysis on important words to demonstrate some potential improvements. We also extend these existing constraints to the semantic versions to better fit the deep learning scenario.

\subsection{Interpretation Methods}\label{sec:IG}
Recently, interpretable machine learning has attracted increasing attention, and many interpretation methods have been proposed, including feature visualization \citep{olah2017feature}, attribution methods\citep{ancona2017unified,baehrens2010explain,shrikumar2017learning,binder2016layer,springenberg2014striving}  and sample importance methods\citep{koh2017understanding}. Among these methods, attribution methods is the most popular approach. It adopts the attribution concept to understand the input output behaviour of a deep neural network. Formally, we have a deep network $F$ with the input $\mathbf{x}=[x_1,...,x_m]$ and ouput $\mathbf{y}=[y_1,...,y_n]$, where $m$ and $n$ separately stands for the dimensions of $x$ and $y$. The goal of attribution methods is to calculate the attrition $A_i=[a_i,...,a_n]$ for each feature of the input $\mathbf{x}$ for the corresponding output value $y_i$. 

Saliency\citep{Simonyan2013Deep} is the first attribution method, it uses gradients to generate the saliency maps. For a given image and the corresponding class saliency map, it first computes the object segmentation mask using the GraphCut\citep{boykov2001interactive} colour segmentation, then calculates the absolute value of $\frac{\partial F(x)}{\partial x_i}$ as the attrition value. Intuitively, this value indicates those input features that can be perturbed the least in order for the target output to change the most. In order to addresses the limitation of gradient-based approaches because the difference from the reference may be non-zero even when the gradient is zero, GradInput \citep{shrikumar2016not} has been proposed. Since GradInput scores are computed using a backpropagation like algorithm, they can be obtained efficiently in a single backward pass after a prediction has been made.
%\begin{equation*}\label{eq:gradinput}
%GradInput(x;F)_i = \left| \frac{\partial F(x)}{\partial x_i}\right|\cdot %x_i.
%\end{equation*}

%Based on two axioms, i.e., \textit{Sensitivity}\citep{friedman2004paths} and \textit{Implementation Invariance} that every attribution method oughts to satisfy. Integrated Gradient is proposed. \textit{Sensitivity}, i.e.,  every input and baseline have different predictions then the differing feature ought to given a non-zero attribution. For  \textit{Implementation Invariance}, i.e., functionally equivalent networks should have the identical attributions. 

Integrated Gradient calculates the average value  of gradients at all points which along a straight line path from the baseline $x^{\prime}$ to input $x$. For image networks, the baseline is the black image \citep{Baehrens2012How}. For text models, the baseline $x^{\prime}$ is set to be zero vector. For the input $x$ and baseline $x^{\prime}$ can be defined as follows which along the $i^{th}$  dimension. Here, $\frac{\partial F(x)}{\partial x_i}$ is the gradient of $F(x)$ along $i^{th}$  dimension.
\begin{equation*}\label{eq:ig}
\begin{split}
IntegratedGrads(x;F)_i &= (x_i -x_i^{\prime})\\ 
& \times \int_{\alpha=0}^{1}\frac{\partial F(x^\prime + \alpha \times (x-x^\prime))}{\partial x_i}\, d\alpha
\end{split}
\end{equation*}
The above formula is the ideal state, but it is hard to calculate. So Integratd Gradient usually adopts the summation operation to approximate the integral operation. To calculate the integral of integrated gradients, we simply summarize the gradients at points along the path from baseline $x^{\prime}$ to input $x$ with the small intervals.

\begin{equation*}\label{eq:ig_approve}
\begin{split}
IntegratedGrads^{approx}(x;F)_i &= \frac{x_i -x_i^{\prime} }{m} \\ 
&\times \sum_{k=1}^m \frac{\partial( F(x^{\prime}+k/m \times (x_i-x_i^{\prime}))}{\partial x_i},
\end{split}
\end{equation*}
where $m$ is the number of steps from  baseline $x^{\prime}$ to input $x$. In theory, the smaller the $m$ is, the closer the two formulas are to each other. we set $m$ to 50  in following experiments. There are also many other paths that monotonically interpolate between baseline $x^{\prime}$ and input $x$. Integrated Gradient has been widely used in interpretating different machine learning methods in text or image applications. Considering the advantage of Integrated Gradient, we use it as our interpretation method to facilitate our study. 
%\begin{figure}
%\setlength{\abovecaptionskip}{0.2cm}
%\setlength{\belowcaptionskip}{-0.2cm}
%	\centering  
%	\includegraphics[width=1.0\linewidth,trim=120 250 350 115,clip]{ig_path}
%	\caption{Three paths between a baseline  $x^{\prime}$ and a input $x$. The path $P_2$ corresponds to the %integrated gradient method, The other two paths stands for other different attribution methods.} 
%	\label{fig:ig_path}   
%\end{figure}

\section{Experiments on IR Heuristics}

In this section, we study interpretation of deep text matching models on IR heuristics. First, we introduce empirical settings, incuding the details of two datasets and the investigated deep text matching models. Then we will describe our interpretation results of these models by using integrated gradient algorithm on four IR heuristics.

\subsection{Empirical Settings}
\subsubsection{Datasets}
To facilitate our empirical study, We experiment  on two datasets, i.e  LETOR4.0 [LT]\footnote{https://www.microsoft.com/en-us/research/publication/letor-benchmark-collection-research-learning-rank-information-retrieval/} and MS Marco[MS] \footnote{http://www.msmarco.org/dataset.aspx}. They are both web search ranking dataset that includes queries and documents. Text matching models can be used to achieve the document ranking list for a specific query. We experiment on both datasets to compare the ranking performances of different models.

LETOR4.0 \citep{qin2013introducing}  is a benchmark data for evaluating learning to rank methods. This dataset sampled from the GOV2 corpus using the TREC 2007 and TREC 2008 to generate two separate subsets,  i.e. MQ2007 and MQ2008. MQ2007 is a bit larger, which contains 1692 queries and 65,323 documents. While MQ2008 only contains 784 queries and 14,384 documents. The query number in MQ2008 is too mall that may cause the serious insufficient training problem, we merge them into one dataset, denoted as LETOR4.0. In total, LETOR4.0 contains 69,623 and 84,834 query-document pairs. The ground-truth labels are collected by human annotators using 3-level graded labeling strategy, i.e. 0, 1, and 2 stands for irrelevant, relevant, and most relevant, respectively.

MS MARCO\citep{nguyen2016ms} is a large scale dataset focused on machine reading comprehension, question answering, and passage ranking. The data are collected from real search engine. All 13000 queries are sampled from real anonymous user queries. The 204638 context passages are extracted from real Web documents. We experiment on the data for passage ranking task. For this task, given a query $q$ and the 1000 candidate passages $P$ = $p1$, $p2$, $p3$,... $p1000$, it is expected that the most relevant passages be ranked as high as possible. Since there are only one document labeled as relevant, the positive and negative data are extremely imbalanced, i.e,, 1000. So we randomly sampled 20 passages from the irrelevant passages to construct our negative samples for each query. In total, 10000, 3000 and 3000 queries are randomly selected to construct the training, validation, and test data, respectively.
%\begin{table}
%\setlength{\abovecaptionskip}{0.0cm}
%\setlength{\belowcaptionskip}{-0.0cm} 
%    \caption{Detailed information of LETOR4.0 and MS Marco datasets.}
%    \label{tab:letor}
%    \centering
%    \begin{tabular}{ccccc}
%        \hline
%         Dataset    &\#Query    &\#Doc  &\#Relevance \\
%        %\cline{2-9}% partial hline from column i to column j
%        \hline
%        \hline
%        LETOR4.0 & 1501 & 57899 & 61480\\
%        MS Marco &13000 & 204638 &215138\\
%        %MS Marco & 198298 & 8494 & 8346\\
%        \hline
%    \end{tabular}
%\end{table} 
%\vspace{-0.2cm} 
\subsubsection{Deep Text Matching Models}
We study both representation and interaction based deep text matching models, and also the hybrid ones. Specifically, ARC-I is chosen as the representative of the representation based models, MatchPyramid, BERT and KNRM are chosen as the representative of the interaction based models. DUET and RI-Match are the hybrid models used in our experiments.

\textbf{ARC-I} utilizes CNN to obtain representations of the input query and document. Then two vectors are concatenated to one vector, and a muti-layer perceptron (MLP) \citep{lin2013network} is used to output the matching score. It concatenates two vectors into one vector. The model is an end-to-end neural network structure \citep{Floyd1999Promoting}.

\textbf{MatchPyramid} [MP] constructs a word level interaction matrix, with each element stands for the similarity of two corresponding words in the query and document. Then interaction matrix is fed as a image to a two dimensional CNN to extract high level matching patterns. Finally, a MLP is used to obtain matching degree.

\textbf{KNRM} uses atching matrix as used in MatchPyramid to obtain the word level matching signals. The difference lies in the second step, where KNRM uses a new kernel-pooling technique, instead of CNN to extract high level matching patterns. The advantage of using the kernel-pooling technique is that they can help to extract multi-level soft match features. At last, a learning-to-rank layer is utilized to combine these features to obtain the final ranking score.

\textbf{DUET} composed of two separate deep neural networks. One matches the query and the document using a local representation. Another one matches the query and the document using learned distributed representations. The two networks are jointly trained as part of a single neural network.

\textbf{RI-Match} combines the benefits of representation and interaction based models. Firstly, the word level and sentence level matching matrices are created by using various matching functions. Then these matrices are fed into a spatial recurrent neural network \citep{wan2016match} to generate high level matching patterns. After a $k$-max pooling \citep{wan2016deep}, the vector is fed into a MLP to output the matching score. 

\textbf{BERT} is a language representation model which stands for Bidirectional Encoder Representations from Transformers. It pre-trains deep bidirectional representation from huge unlabeled text to obtain contextual word representations. The pre-trained BERT model can be further fine-tuned with additional output layer for a specific task. For text matching task, we output the matching degree of two texts as a classification task.

\subsubsection{Parameter Setting}
For all deep models, We trained them by using their implementations in MatchZoo\footnote{https://github.com/NTMC-Community/MatchZoo/tree/1.0}\citep{fan2017matchzoo}. All the hyparameters were tuned using the same experimental setup as described in the respective papers. For the input word embeddings, we initialize the embedding layer with the 300-dimensional Glove\citep{pennington2014glove} word vectors pre-trained in the 840B Common Crawl corpus\footnote{http://index.commoncrawl.org/}. For the out-of-vocabulary (OOV) words, we initialize the word vectors to zero. We leverage Adam\citep{kingma2014adam} as our optimizer to update the parameters of models, and minimize the categorical cross entropy on the training set until the model converges.

\subsubsection{Ranking Performance}
To conduct the interpretation analysis, we need to guarantee that the models have been trained sufficiently. So we first give the ranking performance of the deep text matching on both datasets, as shown in Table \ref{tab:letor_performance} and Table \ref{tab:marco_performance}. From the table, we can see that most deep text matching models have been trained to achieve the SOTA results, except for BERT on LETOR dataset. It is mainly because the dataset size is relatively too small for the huge BERT model and may cause overfitting. Therefore, it is reasonable to conduct further interpretation analysis based on these models.
\begin{table}
    \caption{Performance on Letor4.0 datasets.}
    \label{tab:letor_performance}
    \centering
    \begin{tabular}{ccccc}
        \hline
        Model & MAP(\%) & NDCG@3(\%) & NDCG@5(\%) \\
        %\cline{2-9}% partial hline from column i to column j
        \hline
        ARC-I 	 &42.69		&33.22	&35.28\\
        DUET 	 &43.27		&35.47	&36.98\\
        RI-Match &44.54    	&36.49 	&37.54\\
        MatchPyramid 	&44.37  &36.29	&37.51\\
        KNRM 	&44.06  &36.73	&37.50\\
        BERT    &41.42  &32.42  &   34.46\\
        \hline
    \end{tabular}
\end{table}
\begin{table} 
    \caption{Performance on MS Marco datasets.}
    \label{tab:marco_performance}
    \centering
    \begin{tabular}{ccccc}
        \hline
        Model & MRR(\%) & NDCG@3(\%) & NDCG@5(\%) \\
        %\cline{2-9}% partial hline from column i to column j
        \hline
        ARC-I 	 &50.06		&49.99	&54.13\\ 
        DUET 	&50.70		&50.15	&54.10\\
        RI-Match &52.21		&51.86	&55.77\\
        MatchPyramid 	&52.57		&51.94	&55.49\\
        KNRM 	&52.35		&50.77	&55.79\\
        BERT    &55.62      &54.38  &55.16\\
        \hline   
    \end{tabular}
    \vspace{-5 pt}
\end{table}

\begin{comment}
\begin{figure}
	\centering  
	\includegraphics[width=1.0\linewidth,trim=10 20 10 10,clip]{ig_arci}
	\caption{Attributions from ARC-I model. Term color indicates attribution strength--Red is positive, Blue is negative, and white is neutral.} 
	\label{fig:ig_arci}   
\end{figure}
\end{comment}

\begin{figure}
	\centering  
	\includegraphics[width=0.95\linewidth]{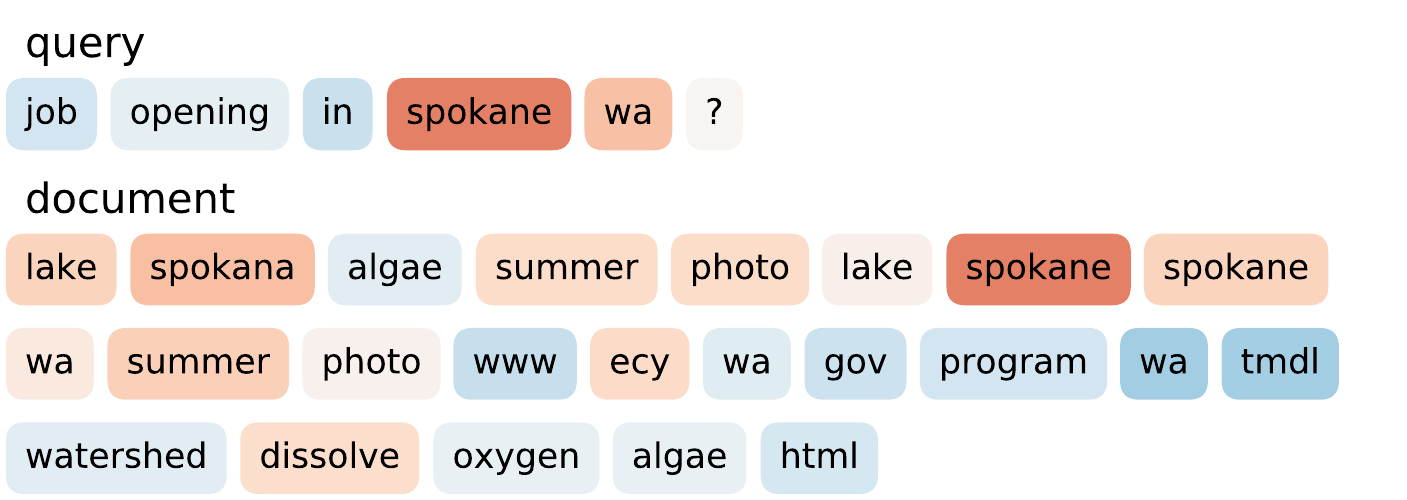}
	\caption{Attribution results for ARC-I model. The color of each term indicates the attribution value, where red is positive, blue is negative, and white is neutral.} 
	\label{fig:ig_arci}  
	\vspace{-5 pt}
\end{figure}

\subsection{Interpretation Analysis}
In this paper, we use Integrated Gradient as the interpretation method to analyze the deep text matching models. As we introduced in Section~\ref{sec:IG}, it computes the integral of integrated gradients to show the importance of each input attribution for the output. Applying IG to our analysis, we can view each trained deep text matching model as the function $F$ in the computation of IG, and output the integral of integrated gradients. For visualization, we use the brightness of different colors to show the value of these gradients. Therefore, we can obtain the significance of each word both in query and document, to show their contributions to the matching score. Figure \ref{fig:ig_arci} shows an example of such analysis. From this example, we can see that the word "\textit{spokane}" is the most attributed term to the matching score of the example query and document, which is accordant with human's understanding. In the following experiments, we will continue to use this analysis technique to facilitate our study.

Before we begin our analysis on IR intrinsics, we first introduce some notations. Formally, we use $q=(q_1,\cdots,q_m)$ to denote a query, $d$ or $d_i$ to denote a document, $\omega$ or $\omega_i$ to denote a query term, and $\omega\prime$ to denote a non-query term. The length of document $d$ is expressed as $|d|$. $c(\omega,d)$ stands for the count of word $\omega$ in document $d$. $f$ stands for a matching function, and $f(d,q)$ calculates the matching score of document $d$ with respect to query $q$. $idf(\omega)$ stands for the IDF discrimination value of a query term $\omega$. While $df(\omega)$ and $tf(\omega)$ denote the term frequency of term $\omega$ in the datasets and the document, respectively.

Now we formally study whether the above learned deep text matching models satisfy the four IR intrinsic constraints,  i.e., term frequency constraint(TFCs), term discrimination constraint (TDC), length normaliza-tion constraints (LNCs), and TF-length constraint (TF-LNC). We first introduce the detailed definition of each constraint, and then demonstrate how we construct data to test whether the trained deep text matching models satisfy the constraints. We also show some further investigations on the reason of the results.

\subsection{Term Frequency Constraint}
There are two term frequency constraint, denoted as TFC1 and TFC2. Both constraints are to capture the desired contribution
of the TF of a term to scoring. The first constraint captures
the basic TF heuristic, which gives a higher score to a document with more occurrences of a query term when the only
difference between two documents is the occurrences of the
query term. While the second constraint ensures that the increase in
the score due to an increase in TF is smaller for larger TFs
(i.e., the second partial derivative w.r.t. the TF variable
should be negative). The formal definitions are show as follows.

\textbf{ TFC1:} Let $q={\{\omega\}}$ which has only one term $\omega$. Assume $|d_1|=|d_2|$, if $c(\omega,d_1)>c(\omega,d_2)$, then$f(d_1,q)>f(d_2,q)$.

\textbf{ TFC2:} Let $q={\{\omega\}}$ which has only one term $\omega$. Assume $|d_1|=|d_2|= |d_3|$, $c(\omega ,d_1)>0$, if $c(\omega,d_2)-c(\omega,d_1)=1$ and $c(\omega,d_3)-c(\omega,d_2)=1$, then $f(d_2,q)-f(d_1,q)>f(d_3,q)-f(d_2,q)$.

To evaluate how much the learned matching function satisfy the desired TFC constraints, we need to construct data which satisfy the above conditions. For TFC1, we can see that the condition is mainly on the query and document length, so we can construct data as follows. Suppose the query $q$ contains $m$ query terms $\{q_1,...,q_m\}$. For each two associated documents $d_1 $ and $d_2$, we can truncate them to be with length $min (| d_1 |, | d_2 |)$, still denoted as $d_1 $ and $d_2$. Then each $q_i, d1$ and $q_i, d_2$ becomes a pair satisfying the condition of TFC1, we can test whether the learned function output an accordant score w.r.t. the occurrence of the query term in each document.

The data construction for TFC2 is a little bit more complicated. For query term $\omega$, we first select three documents that contains $\omega$. Then we select three documents according to the occurrence of $\omega$ in the documents. The document with least $\omega$ is denoted as $d_1$. For $d_2$, we delete extra $\omega$ to make $c(\omega,d_2)-c(\omega,d_1)=1$. If   $c(\omega,d_2)=c(\omega,d_1)$, we add one $\omega$ to the $d_2$ randomly. Then we need to make $|d_1|=|d_2|$. If $|d_1|<|d_2|$, we delete other words in $d_2$ except for $\omega$ until $|d_1|=|d_2|$. If $|d_1|>|d_2|$, we add other words to $d_2$ except for $\omega$ until $|d_1|=|d_2|$. For $d_3$, we do similar constructions to make $|d_2|=|d_3|$ and $c(\omega,d_2)-c(\omega,d_1)=1$.

To evaluate the degree to which the deep text matching models satisfy the TFC constraints, we calculate the proportion of data where the constraints are satisfied. Please note that the data construction could be conducted on both training and test data for LETOR4.0 and Ms MARCO, so we give the experimental results on those four data, denoted as LT-Train, LT-Test, MS-Train, and MS-Test, respectively, as shown in Table \ref{tab:tfc1} and Table \ref{tab:tfc2}.
%\begin{equation}\label{eq:evaluation}
%Accuarcy = \frac{SatisfiedNum}{DataNum}
%\end{equation}
%where $DataNum$ is the number of dataset we construct for the corresponding %constraint, and $SatisfiedNum$ is the satisfied number of $DataNum$. If accuracy %is higher, the constraint is better satisfied in deep text matching models. 

\begin{table}
    \caption{Results of deep text matching models on TFC1.}
    \label{tab:tfc1}
    \centering
    \begin{tabular}{ccccc}
        \hline
        Models & LT-Train(\%)& LT-Test(\%) &MS-Train(\%)&MS-Test(\%) \\
        %\cline{2-9}% partial hline from column i to column j
        \hline
        ARC-I 	    &79.96	&	74.54&	87.17   &84.89	\\
        DUET 	    &80.36 	&	75.37&	88.27   &85.97	\\
        RI-Match 	&81.61	&	76.98&	90.82   &87.53\\
        MP          &93.95	&	81.61&	95.66   &91.54\\
        KNRM 	    &95.68	&	89.23&	94.84   &90.37	\\
        BERT        &77.36  &   75.28&  96.57   &92.46  \\
        \hline
    \end{tabular}
\end{table}
%\begin{figure}
	%\centering  
	%\includegraphics[width=1.0\linewidth]{ig_tfc1}
	%\includegraphics[width=0.9\linewidth]%{Img/F2.pdf}
	%\caption{Attributions result from %MatchPyramid for TFC1.} 
	%\label{fig:ig_tfc1}   
%\end{figure}

\begin{table}
    \caption{Results of deep text matching models on TFC1 with $df(\omega) < 5000$.}
    \label{tab:tfc1_cons}
    \centering
    \begin{tabular}{ccccc}
        \hline
        Models & LT-Train(\%)& LT-Test(\%) &MS-Train(\%)&MS-Test(\%)  \\
        %\cline{2-9}% partial hline from column i to column j
        \hline
        ARC-I 	    &81.47  &77.23	 &88.87  &82.82\\
        DUET 	    &82.63 	&78.46   &89.92	 &82.66\\
        RI-Match 	&83.56	&78.64   &92.62  &86.89\\
        MP          &95.06	&87.01   &95.99  &88.33\\
        KNRM 	    &96.20	&88.52   &95.23  &87.71\\
        BERT        &79.54  &77.12   &97.47  &90.45\\
        \hline
    \end{tabular}
\end{table}

\begin{table}
    \caption{Results of deep text matching models on TFC2.}
    \label{tab:tfc2}
    \centering
    \begin{tabular}{ccccc}
        \hline
        Models & LT-Train(\%)& LT-Test(\%) &MS-Train(\%)&MS-Test(\%) \\
        %\cline{2-9}% partial hline from column i to column j
        \hline
        ARC-I 	    &83.44	&77.35  &	82.83   &78.64	\\
        DUET 	    &84.84 	&78.29  &	82.20   &79.74	\\
        RI-Match 	&85.62	&78.40  &	84.43   &79.28\\
        MP          &87.27	&79.46  &	86.26   &82.68\\
        KNRM 	    &86.92	&78.82  &	87.23   &84.67	\\
        BERT        &81.11  &79.27  &   90.89   &88.56    \\
        \hline
    \end{tabular}
\end{table}

From Table 4, we can see that all deep text matching models satisfy the TFC constraints with a high probability. TFC2 result is not as good as TFC1 result. That is mainly because a lot of matching degree attributes to some more frequent words, such as "\textit{in}" and "\textit{for}", shown as in Fig.~\ref{fig:ig_arci}. As stated in \cite{fang2004formal, fang2005exploration}, words with large DF usually play a negative correlation role in the matching process, so we limit the $df$ of all words to eliminate the influence of these words. Table \ref{tab:tfc1_cons} show the performance of different deep text matching models in terms of TFC1 under the condition $df(\omega) < 5000$ in the training data, where consistency is significantly improved. 

\subsection{Term Discrimination Constraint}
Term Discrimination Constraint captures the interaction between TF and IDF, and emphasizes the effect of using IDF in the scoring of text matching, denoted as TDC. Specifically, given a fixed times of occurrences of query terms, a document should obtain higher matching score if it has more discriminative terms, measured by IDF. The formal definition is shown as follows.
\textbf{TDC:} Let $q$ be a query and has two query terms, then $q={\{\omega_1,\omega_2\}}$. Assume $|d_1|=|d_2|$, $c(\omega_1,d_1)+c(\omega_2,d_1)=c(\omega_1,d_2)+c(\omega_2,d_2)$. If $idf(\omega_1) \ge idf(\omega_2)$ and $c(\omega_1,d_1)>c(\omega_1,d_2)$, then $f(d_1,q)>f(d_2,q)$.

To evaluate how much the learned matching function satisfy the desired TDC constraints, we need to construct data which satisfy the above conditions. Suppose the query $q$ contains several words $\{q_1,....q_m\}$ and two associated documents $d_1$ and $d_2$. We select two words appeared both in $d_1 $ and $d_2$ to construct a new query $q=\{\omega_1,\omega_2\}$. The occurrences of the two words in the document are marked as $c(\omega_1,\omega_2;d_1)$ and $c(\omega_1,\omega_2;d_2)$. Without loss of generality, we set $c(\omega_1,\omega_2 ;d_1)>c(\omega_1,\omega_2;d_2)$, and delete $\omega_1$ or $\omega_2$ in $d_1$ to  make $c(\omega_1,\omega_2;d_1)=c(\omega_1 , \omega_2;d_2)$. Then we delete other words to make the two documents with equal length, i.e.,~$|d_1|=|d_2|$. 

To evaluate the degree to which the deep text matching models satisfy the TDC constraint, we calculate the proportion of data where the constraint is satisfied. The results on LT-Train, LT-Test, MS-Train, and MS-Test are shown in Table~\ref{tab:tdc} and Table \ref{tab:tdc_yizhi}. For the deep text matching models, they all satisfy the TDC results with a high probability in statistics. In \citep{fang2004formal, fang2005exploration}, a stronger condition is added to TDC constraint, that is $c(\omega_1, d_2) \leq c(\omega_2, d_1)$. So we also investigate the influence of this condition for interpretaing existing deep text matching models. Specifically, with this condition of TDC, the proportion of data which satisfy TDC is shown in Table \ref {tab:tdc_yizhi}. For the experimental results, we can see that deep text matching better satisfy TDC. That is because when $c(\omega_1,d_2) \leq c(\omega_2,d_1)$, the influence of word $ \omega_2$ will be reduced, which makes the influence of $\omega_1$ with high IDF more prominent.

\begin{table}
    \caption{Results of deep text matching models on TDC.}
    \label{tab:tdc}
    \centering
    \begin{tabular}{ccccc}
        \hline
        Models & LT-Train(\%)& LT-Test(\%) &MS-Train(\%)&MS-Test(\%) \\
        %\cline{2-9}% partial hline from column i to column j
        \hline
        ARC-I 	    &83.22  &77.26  &84.70	&81.59	\\
        DUET 	    &84.28	&79.28  &83.11  &79.77	\\
        RI-Match 	&85.28	&79.47  &84.67  &80.54\\
        MP          &86.37	&80.27  &85.68  &81.81\\
        KNRM 	    &85.26	&79.23  &87.89  &83.87\\
        BERT        &79.11  &78.12  &88.23  &85.25\\
        \hline
    \end{tabular}
\end{table}

\begin{table}
    \caption{Results of deep text matching models on TDC with $c(\omega_1,d_2)\leq c(\omega_2,d_1)$.}
    \label{tab:tdc_yizhi}
    \centering
    \begin{tabular}{ccccc}
         \hline
        Models & LT-Train(\%)& LT-Test(\%) &MS-Train(\%)&MS-Test(\%) \\
        %\cline{2-9}% partial hline from column i to column j
        \hline
        ARC-I 	     &87.48    &79.49 &87.83 &83.00\\
        DUET 	     &87.13    &82.23 &86.49 &82.56\\
        RI-Match 	 &88.58    &83.56 &87.76 &84.85\\
        MP           &89.34    &83.65 &88.73 &84.23\\
        KNRM 	     &88.75    &84.68 &90.45 &84.80\\
        BERT         &83.45    &80.45 &91.54 &86.27\\
        \hline
    \end{tabular}
\end{table}
\begin{table}
    \caption{Results of deep text matching models on LNC1.}
    \label{tab:lnc1}
    \centering
    \begin{tabular}{ccccc}
         \hline
        Models & LT-Train(\%)& LT-Test(\%) &MS-Train(\%)&MS-Test(\%) \\
        %\cline{2-9}% partial hline from column i to column j
        \hline
        ARC-I 	     &70.36 &67.23  &70.21 &66.67\\
        DUET 	     &71.55 &67.94  &72.63 &68.89\\
        RI-Match 	 &72.56 &68.34  &71.35 &68.12\\
        MP           &74.28 &69.45  &72.66 &67.28\\
        KNRM 	     &74.18 &69.26  &71.38 &68.31\\
        BERT         &69.24 &66.26  &73.57 &69.23\\
        \hline
    \end{tabular}
\end{table}

\begin{table}
    \caption{Results of deep text matching models on LNC2.}
    \label{tab:lnc2}
    \centering
    \begin{tabular}{ccccc}
         \hline
        Models & LT-Train(\%)& LT-Test(\%) &MS-Train(\%)&MS-Test(\%) \\
        %\cline{2-9}% partial hline from column i to column j
        \hline
        ARC-I 	     &96.7  &88.12  &87.23   &83.59\\
        DUET 	     &96.23 &87.93  &88.25   &84.38\\
        RI-Match 	 &96.56 &87.85  &89.35   &84.67\\
        MP           &100.00&89.87  &96.07   &90.28\\
        KNRM 	     &99.01 &87.28  &93.79  &89.06\\
        BERT         &94.34 &986.23 &100.00 &93.26\\
        \hline
    \end{tabular}
\end{table}
\subsection{Length Normalization Constraint}
There are two length normalization constraints, denoted as LNC1 and LNC2. Both constraints capture contribution of the length of document in the scoring process. LNC1 says that if we add one extra non-relevant word to form a new document, then the matching degree of the new document with respect to the query will decrease. While LNC2 says that if we duplicate a document $k$ times to form a new document, the new document will obtain higher matching score than the original document. The formal definitions are shown as follows.

\textbf{LNC1:} Let $q$ be a query and $d_1$, $d_2$ be two document. If for some word $\omega^\prime \not\in q$, $c(\omega^\prime,d_2)=c(\omega^\prime,d_1)+1$, but for any query term $\omega$, $c(\omega,d_2)=c(\omega,d_1)$, we have $f(d_1,q)>f(d_2,q)$.

\textbf{LNC2:} Let $q$ be a query. $\forall k >1$, $d_1$ and $d_2$ are two documents with $|d_1| = k \cdot |d_2|$. If for any query term $\omega$, $c(\omega,d_1) = k\cdot c(\omega,d_2)$, we have $f(d_1,q)>f(d_2,q)$.

To evaluate how much the learned matching function satisfy the desired LNC constraints, we need to construct data which satisfy the constrains. For LNC1, suppose that the query $q$ contains several terms $\{q_1,....q_m\}$ and the document is $d_1$. We first find a word in the document that does not exist in the query $q$. Appending this word to the end of the document $d_1 $ to form a new document $d_2$, then $(q,d)$ and $(q, d_2)$ form a data pair that satisfies the LNC1 constraint. 

The data construction for LNC2 is a little bit more easy. Suppose that the query is $q$ and the document is $d_1 $ with length $|d_1|$. Here, we set $k = 2$ as an example. We first duplicate document $d_1$ to form the new document $d_2$ with length $2|d_1|$, then $(q, d_1)$ and $(q, d_2)$ forms a data pair that satisfies the constraints. 

After that, we calculate the proportion of data that satisfy the constraints. The results on  LT-Train, LT-Test, MS-Train, and MS-Test are shown in Table \ref{tab:lnc1} and Table \ref{tab:lnc2}. From the results, we can see that LNC1 constraint is not so well satisfied for deep text matching models as LNC2. So we utilize the IG algorithm to conduct the attribution analyse. 

\begin{figure}
	\centering  
	\includegraphics[width=0.9\linewidth]{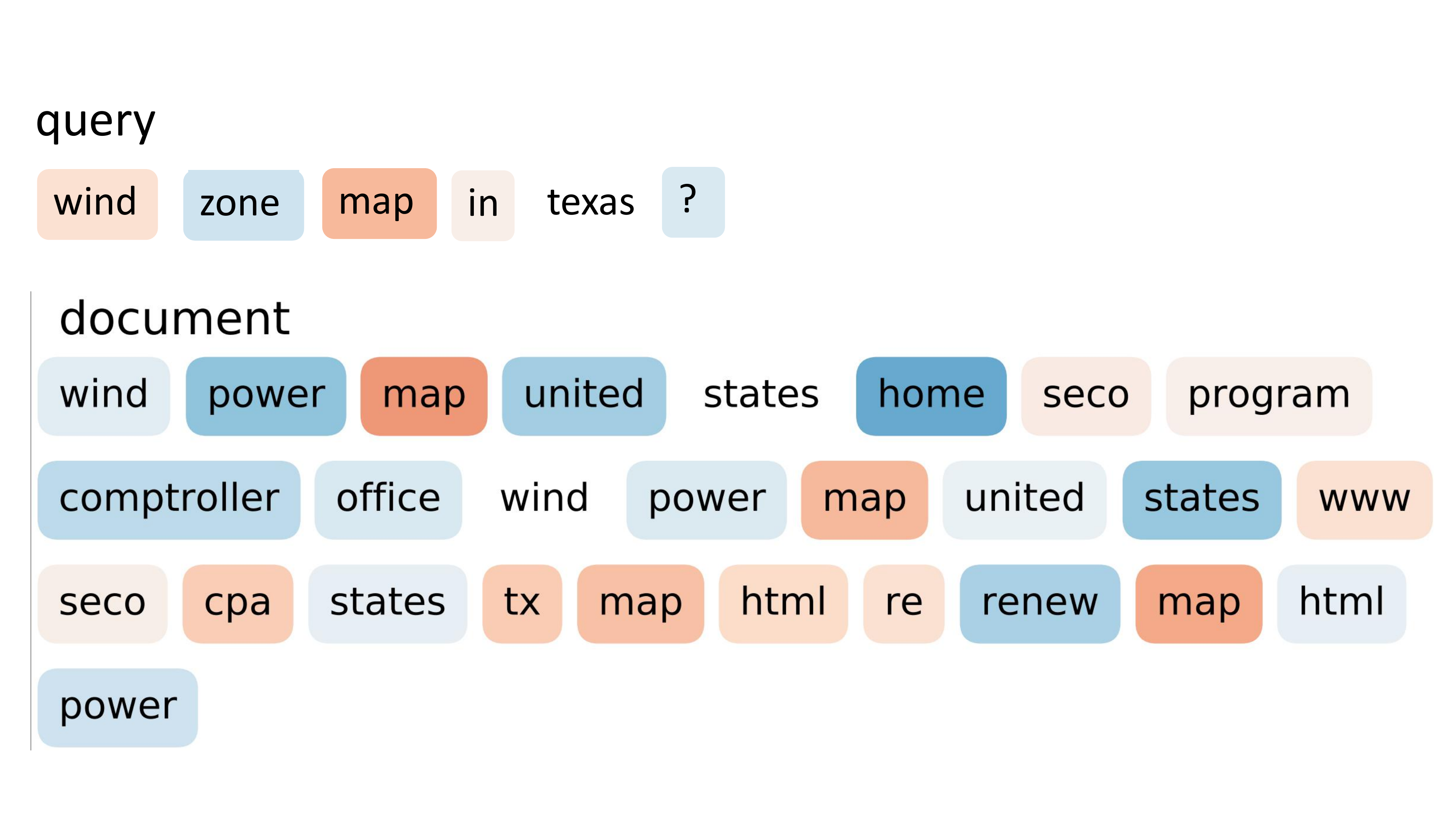}
	\caption{An example for LNC1.} 
	\label{fig:lnc1_neg}   
\end{figure}

\begin{figure}
	\centering  
	\includegraphics[width=0.9\linewidth]{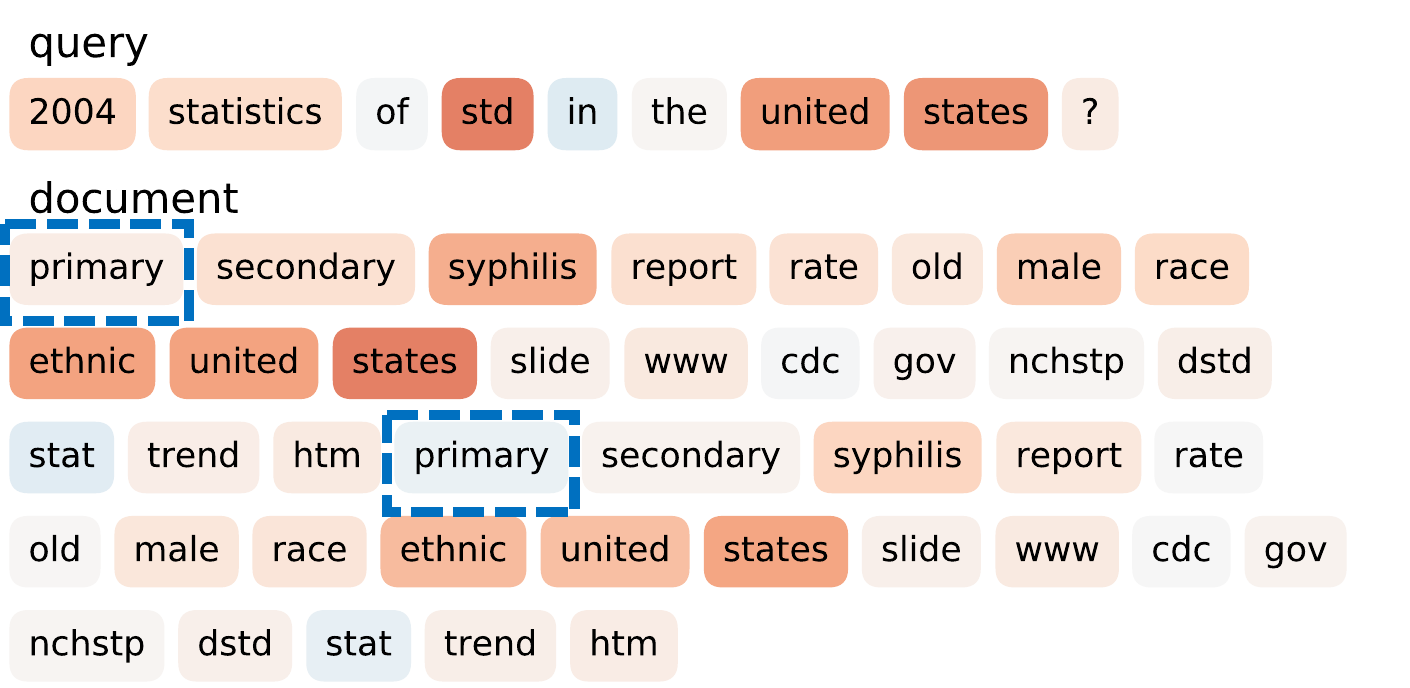}
	\caption{An example for LNC2.} 
	\label{fig:ig_lnc2}   
\end{figure}

We found that one key difference between LNC1 and LNC2 is that, the influence of duplicated words are different. We show two examples in Figure \ref{fig:lnc1_neg} and \ref{fig:ig_lnc2}. We can see that the word "\textit{map}" has the positive attribution value in the original document. When it is added to form a new document, it still has a positive attribution value and will improve the matching degree of the documents. That is contradiction with the LNC1 constraint. While for LNC2, though most duplicated word still attribute with the same sign, some key words like "\textit{primary}" change their attribution sign from positive to negative. So we conclude that the attribution sign plays an important role in LNC1, and we need to take this factor into account. Specifically, when we construct the data for LNC1, the attribution value of the word $\omega\prime$ is constrained to be less than zero. In this way, the proportion that satisfy the new constraint of existing deep text matching models are shown in Table \ref{tab:lnc1_con}. On the contrary, the results are shown in Table \ref{tab:lnc1_con_neg} for adding words with positive attribution value. From the results, we can see that the proportion of data satisfying LNC1 is significantly improved by adding the condition $IG\_value(\omega\prime)>0$.


\begin{table}
    \caption{Results of deep text matching models on LNC1 with $IG\_value(\omega\prime)<0$.}
    \label{tab:lnc1_con}
    \centering
    \begin{tabular}{ccccc}
         \hline
        Models & LT-Train(\%)& LT-Test(\%) &MS-Train(\%)&MS-Test(\%) \\
        %\cline{2-9}% partial hline from column i to column j
        \hline
        ARC-I 	     &83.57 &80.37  &85.99 &82.09\\
        DUET 	     &84.62 &81.88  &86.46 &82.72\\
        RI-Match 	 &86.35 &82.55  &87.86 &83.86\\
        MP           &87.29 &83.54  &87.12 &84.21\\
        KNRM 	     &86.48 &82.70  &88.08 &84.92\\
        BERT         &81.37 &80.03  &90.32 &87.53\\
        \hline
    \end{tabular}
\end{table}
\begin{table}
    \caption{Results of deep text matching models on LNC1 with $IG\_value(\omega\prime)>0$.}
    \label{tab:lnc1_con_neg}
    \centering
    \begin{tabular}{ccccc}
     \hline
        Models & LT-Train(\%)& LT-Test(\%) &MS-Train(\%)&MS-Test(\%) \\
        %\cline{2-9}% partial hline from column i to column j
        \hline
        ARC-I 	     &62.37 &55.32  &61.41 &58.54\\
        DUET 	     &63.56 &58.35  &62.23 &59.47\\
        RI-Match 	 &66.45 &63.26 &62.46 &60.48\\
        MP           &67.57 &64.30  &64.24 &62.67\\
        KNRM 	     &66.25 &60.37  &64.28 &62.25\\
        BERT         &60.23 &51.46  &59.28 &57.28\\
        %\hline
        %\cline{2-9}% partial hline from column i to column j
        \hline
    \end{tabular}
\end{table}

\begin{table}
    \caption{The ratio of duplicated words with consistency attributions on LNC2.}
    \label{tab:lnc2_yizhi}
    \centering
    \begin{tabular}{ccccccc}
        \hline
        DataSet & ARC-I &DUET & RI-Match& MP &KNRM &BERT\\
        %\cline{2-9}% partial hline from column i to column j
        \hline
        \hline
        Letor 4.0 	 &0.707 &0.728	&0.739	&0.758	&0.742 &0.693\\
        MS Marco 	 &0.692 &0.712	&0.724	&0.735	&0.726 &0.758\\
        \hline
    \end{tabular}
\end{table}

As shown in the Table \ref{tab:lnc2}, the deep text matching models well satisfy LNC2. Furthermore, we conduct an experiment to study the influence of duplicated contents. Specifically for each document, we count the proportion of words in the duplicate part whose attribution is weaker than that in the previous part. Here we do not distinguish the training and test data, and just average the results, shown in Table \ref{tab:lnc2_yizhi}. From results, we can see that the a large proportion of words in the duplicate part contribute weaker with the sign than that in the previous part. Considering the fact that the role of most words is consistent with that in the previous part, Although these functions are weaker than those of the previous part, the consistency ratio of these words is helpful to enhance the matching degree between query and document globally. 

\subsection{TF-Length Constraint}
TF-Length constraint captures the interaction between TF and document length, denoted as TF-LNC. It says that if $d_1$ is constructed by adding more query term to $d_2$, the matching score of $d_1$ will be higher than $d_2$. The formal definition is shown as follows.

\textbf{TF-LNC:} Let $q={\{\omega}$ be the query which has only one term $\omega$. Assume $c(\omega,d_1)>c(\omega,d_2)$ and $|d_1|=|d_2| + c(\omega,d_1)-c(\omega,d_2)$, we have $f(d_1,q)>f(d_2,q)$.
\begin{figure}
	\centering  
	\includegraphics[width=0.9\linewidth]{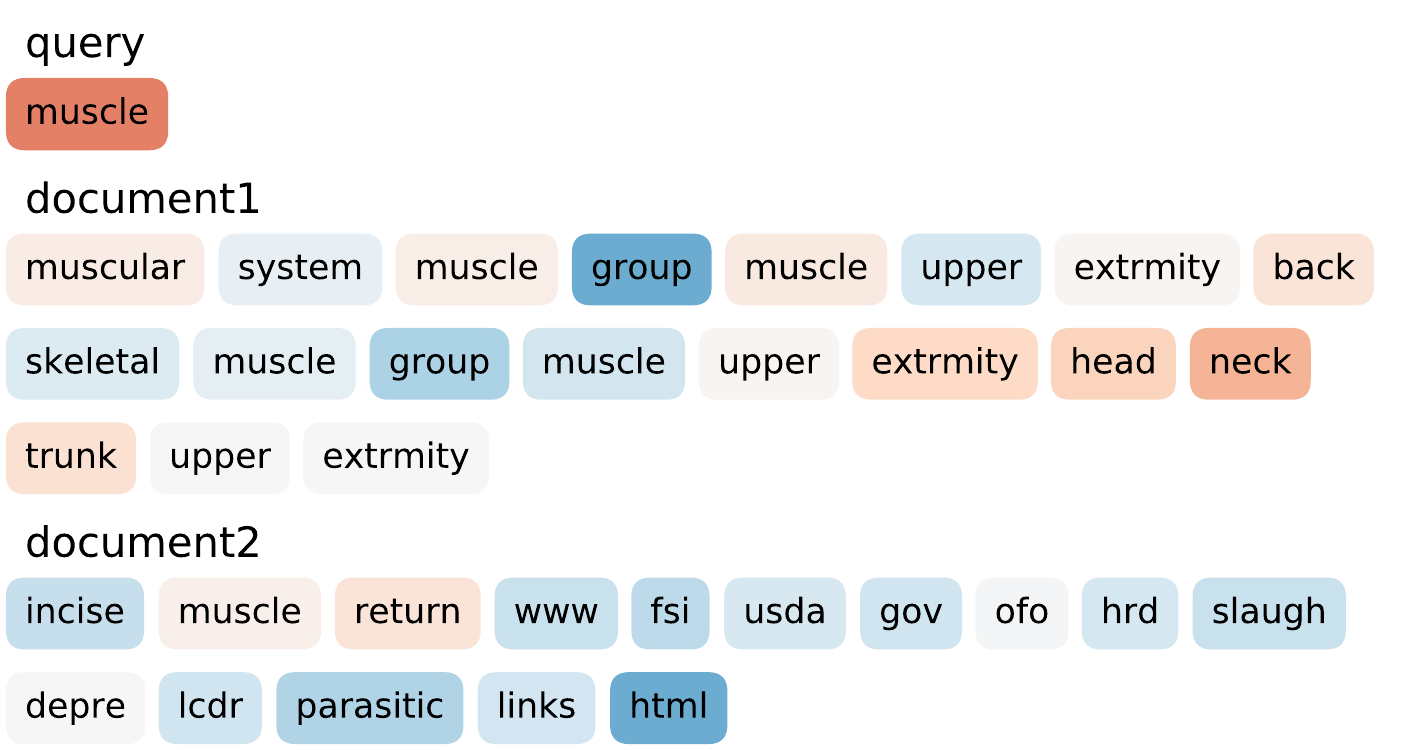}
	\caption{An example for TF-LNC.} 
	\label{fig:ig_tf-lnc}   
\end{figure}
To evaluate how much the learned matching function satisfy the desired TF-LNC constraint, we need to construct data which satisfy the above condition in the definition. We first add $c(\omega, d_1)-c(\omega,d_2)$ words (not $\omega$) to document $d_2$. Then we calculate the proportion of data where the constraint is satisfied, and the results are shown in Table \ref{tab:tf-lnc}. From the table, we can see that most of deep text matching models well satisfy the TF-LNC. We further apply the IG algorithm to analyse the attribution of each word, and an example is shown in Figure \ref{fig:ig_tf-lnc}. In the example, the query contains only one query term "\textit{muscle}", and $c(q, d_1)> c(q, d_2) $. Although the document length of $d_1$ is larger, most of the "\textit{muscle}" appearing in the document attribute positively to the matching score. That explains why $d_1$ is more relevant than $d_2$ with respect to the query. We further make a statistics on the proportion of words in $d_1$ with greater attribution value than $d_2$ in the dataset, as shown in Table \ref{tab:tf-lnc_yizhi}. We can see that most words remain their attribution signs in the duplication process.
\begin{table}
\setlength{\abovecaptionskip}{0.0cm}
\setlength{\belowcaptionskip}{-0.0cm}
    \caption{Results of deep text matching models on TF-LNC.}
    \label{tab:tf-lnc}
    \centering
    \begin{tabular}{ccccc}
         \hline
        Models & LT-Train(\%)& LT-Test(\%) &MS-Train(\%)&MS-Test(\%) \\
        %\cline{2-9}% partial hline from column i to column j
        \hline
        ARC-I 	     &84.37 &80.35  &83.75   &80.38\\
        DUET 	     &83.62 &80.01  &85.18   &82.44\\
        RI-Match 	 &81.38 &78.36   &82.75   &80.26\\
        MP           &94.00 &87.58   &93.90   &87.28\\
        KNRM 	     &96.39 &88.37   &92. 92  &85. 19\\
        BERT         &80.14 &78.25   &95.28   &88.28\\
        \hline
    \end{tabular}
\end{table}
\vspace{-0.0cm} 

\begin{table}
\setlength{\abovecaptionskip}{0.2cm}
\setlength{\belowcaptionskip}{-0.2cm}
    \caption{The ratio of duplicated words with consistency attributions on TF-LNC.}
    \label{tab:tf-lnc_yizhi}
    \centering
    \begin{tabular}{ccccccc}
         \hline
        Dateset & ARC-I &DUET & RI-Match& MP &KNRM &BERT \\
        %\cline{2-9}% partial hline from column i to column j
        \hline
        \hline
        Letor 4.0  	 &0.772 &0.745	&0.712	&0.738	&0.735  &0.764\\
        MS Marco 	 &0.743 &0.727	&0.704	&0.712	&0.709  &0.700\\
        \hline
    \end{tabular}
\end{table}
\vspace{-0.2cm} 
\section{Extension to Semantic IR Heuristics}
From the above studies, we can conclude that the existing deep text matching models well satisfy the four IR heuristics. However, all the IR heuristics only consider the exact matching, which may be limited in the semantic scenario that deep learning models are good at. So we propose to extend the previous IR heuristics to incorporate the semantic meanings, namely semantic IR heuristics including TFC1-E, TDC-E, and TF-LNC-E. The precise definitions are described as follows.

\textbf{TFC1-E:} Let $q={\{\omega\}}$ which has only one term $\omega$. Assume $|d_1|=|d_2|$, $\theta \in [0,1]$ is the threshold of the cosine similarity, $\gamma_{i}$ stands for the $i$-th word in document $d$. We define the semantic count [Sc] of $\omega$ for $d$ in the Equation \ref{eq:sc}. Assume $Sc(\omega,d_1)\geq Sc(\omega,d_2)$, where $f$ denotes the cosine similarity function, we ave $f(d_1,q)>f(d_2,q)$.

\begin{equation}\label{eq:sc}
SemanticCount(\omega,d) = \sum_{i=1}^{|d|} [f(\omega,\gamma_{i})|f(\omega,\gamma_{i})\geq \theta]
\end{equation}

\textbf{TDC-E:} Let $q$ be a query which has two query terms $q={\{\omega_1,\omega_2\}}$. Assume $|d_1|=|d_2|$,  $[Sc(\omega_1,d_1) +  Sc(\omega_2,d_1)]$ $-[Sc(\omega_1, d_2) + Sc(\omega_2, d_2)]$ $< \epsilon $, here we set $\epsilon=0.1$. If $idf(\omega_1) \ge idf(\omega_2)$ and $Sc(\omega_1,d_1)>Sc(\omega_1,d_2)$, we have $f(d_1,q)>f(d_2,q)$.

\textbf{ TF-LNC-E:} Let $q={\{\omega\}}$ which has only one term $\omega$. Assume $|d_1| = |d_2| + \left\lfloor  Sc(\omega, d_1) - Sc(\omega, d_2)\right\rfloor$ and $Sc(\omega, d_1) > Sc(\omega, d_2)$ , where  the $\left\lfloor \right\rfloor$ denotes the floor function, we have $f(d_1, q)>f(d_2, q)$.

We compare the previous IR heuristics and our proposed extension versions by comparing the satisfied data proportion, shown in Table \ref{tab:extention1}, \ref{tab:extention2}, \ref{tab:extention3}, and \ref{tab:extention4}. The experimental results show that existing deep text matching models better satisfy our proposed extension versions than the previous IR heuristics, which better explain the existing deep text matching models than traditional ones.

\begin{table}
    \caption{The results on semantic IR heuristics in the training data of LETOR when $\theta>0.90$.}
    \label{tab:extention1}
    \centering
    \begin{tabular}{ccccccc}
        \hline
        Constraint & ARC-I &DUET & RI-Match &MP &KNRM &BERT \\
        %\cline{2-9}% partial hline from column i to column j
        \hline
        \hline
        TFC1 	 &79.96	&80.36	&81.61&	93.95&95.68 &77.37	\\
        TFC1-E 	 &\textbf{82.67}&\textbf{84.37}	&	\textbf{86.59}&	\textbf{94.02}&\textbf{96.78}&\textbf{80.36}		\\
        \hline
        TDC 	 &83.22 &84.28	&85.28	&86.37 &85.26 &79.11\\
        TDC-E   &\textbf{83.88}    &\textbf{84.36}    &\textbf{85.31} &\textbf{86.98}&    \textbf{86.32}&    \textbf{82.28}\\
        \hline
        TF-LNC 	 &84.37	&83.62  &81.38	&94.00  &96.39 &80.14\\
        TF-LNC-E &\textbf{85.79}&\textbf{85.86}	&\textbf{84.84}	&\textbf{94.23}&\textbf{96.89}&\textbf{83.37}		\\
        \hline
    \end{tabular}
\end{table}

\begin{table}
    \caption{The results on semantic IR heuristics in the test data of LETOR when $\theta>0.90$.}
    \label{tab:extention2}
    \centering
    \begin{tabular}{ccccccc}
        \hline
        Constraint & ARC-I &DUET & RI-Match &MP &KNRM  &BERT\\
        %\cline{2-9}% partial hline from column i to column j
        \hline
        \hline
        TFC1 	 &74.54	&75.37 &76.98&	81.17&89.23 &75.28	\\
        TFC1-E 	 &\textbf{75.75}&\textbf{76.25}	&	\textbf{78.89}&	\textbf{82.13}&\textbf{90.25}	&\textbf{75.89}	\\
        \hline
        TDC 	 &77.26&79.28	&79.47	&80.27&79.23    &78.12\\
        TDC-E 	 &\textbf{77.44}	&\textbf{80.31}  &\textbf{80.36}	&\textbf{82.19} & \textbf{80.33} & \textbf{79.25}	\\
        \hline
        TF-LNC 	 &80.35	&80.01  &78.36	&87.58  &88.37  &78.25\\
        TF-LNC-E &\textbf{81.65}&\textbf{82.15}	&\textbf{81.59}	&\textbf{89.97}&\textbf{91.04}	&\textbf{93.11}	\\
        \hline
    \end{tabular}
\end{table}

\begin{table}
    \caption{The results on semantic IR heuristics in the training data of Marco when $\theta>0.90$.}
    \label{tab:extention3}
    \centering
    \begin{tabular}{ccccccc}
        \hline
        Constraint & ARC-I &DUET & RI-Match &MP &KNRM  &BERT\\
        %\cline{2-9}% partial hline from column i to column j
        \hline
        \hline
        TFC1 	 &87.17	&88.27 &90.82&	95.66&94.84 &96.57	\\
        TFC1-E 	 &\textbf{88.44}&\textbf{90.86}	&\textbf{92.35}&	\textbf{95.83}&\textbf{95.17}&\textbf{97.38}		\\
        \hline
        TDC 	 &84.70&83.11	&84.67	&85.68&87.89   &88.23 \\
        TDC-E 	 &\textbf{85.12}	&\textbf{84.47}  &\textbf{85.86}	&\textbf{86.08}	&\textbf{87.44}	&\textbf{88.95}\\
        \hline
        TF-LNC 	 &83.38	&85.18  &82.75	&93.90  &92.92  &95.28\\
        TF-LNC-E &\textbf{86.91}&\textbf{87.75}	&\textbf{83.28}	&\textbf{94.13}&\textbf{93.56}&\textbf{96.24}		\\
        \hline
    \end{tabular}
\end{table}

\begin{table}
    \caption{The results on semantic IR heuristics in the test data of Marco when $\theta>0.90$.}
    \label{tab:extention4}
    \centering
    \begin{tabular}{ccccccc}
        \hline
        Constraint & ARC-I &DUET & RI-Match &MP &KNRM  &BERT\\
        %\cline{2-9}% partial hline from column i to column j
        \hline
        \hline
        TFC1 	 &84.89	&85.97 &87.97&	91.54&90.37 &92.46	\\
        TFC1-E 	 &\textbf{85.52}&\textbf{87.82}	&	\textbf{88.40}&	\textbf{92.25}&\textbf{91.26}&\textbf{92.75}		\\
        \hline
        TDC 	 &81.59&79.77	&80.54	&81.81&83.87    &85.25\\
        TDC-E 	 &\textbf{81.89}	&\textbf{80.43}  &\textbf{80.98}	&\textbf{82.23}	&\textbf{81.32}	&\textbf{86.13}\\
        \hline
        TF-LNC 	 &80.38	&82.44  &80.26	&87.28  &85.19  &88.28\\
        TF-LNC-E &\textbf{81.88}&\textbf{83.16}	&\textbf{82.34}	&\textbf{89.19}&\textbf{86.04}&\textbf{89.46}		\\
        \hline
    \end{tabular}
\end{table}
	
\section{Conclusion}
In this paper, we propose to understand deep text matching model from the perspective of how much do they satisfy the IR heuristics. We propose an empirical method to facilitate our study. First, we train deep text matching model on original training data, and then apply it to some constructed data satisfying the assumption of a constraint. As a result, the proportion of data satisfying the constraint can be used as our required qualitative measure. In our experiments, we test six representative deep text matching models (ARC-I, MatchPyramid, KNRM, RI-Match, BERT and DUET), in terms of four IR heuristics (TFCs, TDC, LNCs, and TF-LNC). Experimental results show that all six models satisfy heuristics with high probabilities in statistics. Moreover, we extend the existing IR heuristics to the semantic version, and experimental results show that these semantic constraints can be better satisfied by these deep text matching models. So the semantic IR heuristics can better explain the success of deep text matching models, as compared with traditional ones. Except for these revealed understandings, We believe the proposed evaluation methodology will be useful for testing existing and future deep text matching models.

In future, we plan to extend our study to other deep text matching models and IR heuristics, to complete a more thorough investigation. Furthermore, we are interested in how to design more suitable IR heuristics for deep learning, and how to use the proposed semantic heuristics to help us design better deep text matching models.
%\input{samplebody-conf}
%\newpage
\balance
\bibliographystyle{ACM-Reference-Format}
\bibliography{reference}


\begin{thebibliography}{45}

%%% the Web address.

\ifx \showCODEN    \undefined \def \showCODEN     #1{\unskip}     \fi
\ifx \showDOI      \undefined \def \showDOI       #1{#1}\fi
\ifx \showISBNx    \undefined \def \showISBNx     #1{\unskip}     \fi
\ifx \showISBNxiii \undefined \def \showISBNxiii  #1{\unskip}     \fi
\ifx \showISSN     \undefined \def \showISSN      #1{\unskip}     \fi
\ifx \showLCCN     \undefined \def \showLCCN      #1{\unskip}     \fi
\ifx \shownote     \undefined \def \shownote      #1{#1}          \fi
\ifx \showarticletitle \undefined \def \showarticletitle #1{#1}   \fi
\ifx \showURL      \undefined \def \showURL       {\relax}        \fi
% invisible to TeX
\providecommand\bibfield[2]{#2}
\providecommand\bibinfo[2]{#2}
\providecommand\natexlab[1]{#1}
\providecommand\showeprint[2][]{arXiv:#2}

\bibitem[\protect\citeauthoryear{Ancona, Ceolini, {\"O}ztireli, and
  Gross}{Ancona et~al\mbox{.}}{2017}]%
        {ancona2017unified}
\bibfield{author}{\bibinfo{person}{Marco Ancona}, \bibinfo{person}{Enea
  Ceolini}, \bibinfo{person}{Cengiz {\"O}ztireli}, {and}
  \bibinfo{person}{Markus Gross}.} \bibinfo{year}{2017}\natexlab{}.
\newblock \showarticletitle{A unified view of gradient-based attribution
  methods for deep neural networks}. In \bibinfo{booktitle}{\emph{NIPS
  2017-Workshop on Interpreting, Explaining and Visualizing Deep Learning}}.
  ETH Zurich.
\newblock


\bibitem[\protect\citeauthoryear{Baehrens, Schroeter, Harmeling, Kawanabe,
  Hansen, and M{\~A}{\v{z}}ller}{Baehrens et~al\mbox{.}}{2010}]%
        {baehrens2010explain}
\bibfield{author}{\bibinfo{person}{David Baehrens}, \bibinfo{person}{Timon
  Schroeter}, \bibinfo{person}{Stefan Harmeling}, \bibinfo{person}{Motoaki
  Kawanabe}, \bibinfo{person}{Katja Hansen}, {and}
  \bibinfo{person}{Klaus-Robert M{\~A}{\v{z}}ller}.}
  \bibinfo{year}{2010}\natexlab{}.
\newblock \showarticletitle{How to explain individual classification
  decisions}.
\newblock \bibinfo{journal}{\emph{Journal of Machine Learning Research}}
  \bibinfo{volume}{11}, \bibinfo{number}{Jun} (\bibinfo{year}{2010}),
  \bibinfo{pages}{1803--1831}.
\newblock


\bibitem[\protect\citeauthoryear{Baehrens, Schroeter, Harmeling, Kawanabe,
  Hansen, and Müller}{Baehrens et~al\mbox{.}}{2012}]%
        {Baehrens2012How}
\bibfield{author}{\bibinfo{person}{David Baehrens}, \bibinfo{person}{Timon
  Schroeter}, \bibinfo{person}{Stefan Harmeling}, \bibinfo{person}{Motoaki
  Kawanabe}, \bibinfo{person}{Katja Hansen}, {and}
  \bibinfo{person}{Klaus~Robert Müller}.} \bibinfo{year}{2012}\natexlab{}.
\newblock \showarticletitle{How to Explain Individual Classification
  Decisions}.
\newblock \bibinfo{journal}{\emph{Journal of Machine Learning Research}}
  \bibinfo{volume}{11}, \bibinfo{number}{9} (\bibinfo{year}{2012}),
  \bibinfo{pages}{1803--1831}.
\newblock


\bibitem[\protect\citeauthoryear{Berger, Caruana, Cohn, Freitag, and
  Mittal}{Berger et~al\mbox{.}}{2000}]%
        {berger2000bridging}
\bibfield{author}{\bibinfo{person}{Adam Berger}, \bibinfo{person}{Rich
  Caruana}, \bibinfo{person}{David Cohn}, \bibinfo{person}{Dayne Freitag},
  {and} \bibinfo{person}{Vibhu Mittal}.} \bibinfo{year}{2000}\natexlab{}.
\newblock \showarticletitle{Bridging the lexical chasm: statistical approaches
  to answer-finding}. In \bibinfo{booktitle}{\emph{Proceedings of the 23rd
  annual international ACM SIGIR conference on Research and development in
  information retrieval}}. ACM, \bibinfo{pages}{192--199}.
\newblock


\bibitem[\protect\citeauthoryear{Binder, Montavon, Lapuschkin, M{\"u}ller, and
  Samek}{Binder et~al\mbox{.}}{2016}]%
        {binder2016layer}
\bibfield{author}{\bibinfo{person}{Alexander Binder},
  \bibinfo{person}{Gr{\'e}goire Montavon}, \bibinfo{person}{Sebastian
  Lapuschkin}, \bibinfo{person}{Klaus-Robert M{\"u}ller}, {and}
  \bibinfo{person}{Wojciech Samek}.} \bibinfo{year}{2016}\natexlab{}.
\newblock \showarticletitle{Layer-wise relevance propagation for neural
  networks with local renormalization layers}. In
  \bibinfo{booktitle}{\emph{International Conference on Artificial Neural
  Networks}}. Springer, \bibinfo{pages}{63--71}.
\newblock


\bibitem[\protect\citeauthoryear{Boykov and Jolly}{Boykov and Jolly}{2001}]%
        {boykov2001interactive}
\bibfield{author}{\bibinfo{person}{Yuri~Y Boykov} {and} \bibinfo{person}{M-P
  Jolly}.} \bibinfo{year}{2001}\natexlab{}.
\newblock \showarticletitle{Interactive graph cuts for optimal boundary \&
  region segmentation of objects in ND images}. In
  \bibinfo{booktitle}{\emph{Proceedings eighth IEEE international conference on
  computer vision. ICCV 2001}}, Vol.~\bibinfo{volume}{1}. IEEE,
  \bibinfo{pages}{105--112}.
\newblock


\bibitem[\protect\citeauthoryear{Chen, Lan, Pang, Guo, Xu, and Cheng}{Chen
  et~al\mbox{.}}{2018}]%
        {chen2018ri}
\bibfield{author}{\bibinfo{person}{Lijuan Chen}, \bibinfo{person}{Yanyan Lan},
  \bibinfo{person}{Liang Pang}, \bibinfo{person}{Jiafeng Guo},
  \bibinfo{person}{Jun Xu}, {and} \bibinfo{person}{Xueqi Cheng}.}
  \bibinfo{year}{2018}\natexlab{}.
\newblock \showarticletitle{RI-Match: Integrating Both Representations and
  Interactions for Deep Semantic Matching}. In \bibinfo{booktitle}{\emph{Asia
  Information Retrieval Symposium}}. Springer, \bibinfo{pages}{90--102}.
\newblock


\bibitem[\protect\citeauthoryear{Denil, Demiraj, Kalchbrenner, Blunsom, and
  de~Freitas}{Denil et~al\mbox{.}}{2014}]%
        {denil2014modelling}
\bibfield{author}{\bibinfo{person}{Misha Denil}, \bibinfo{person}{Alban
  Demiraj}, \bibinfo{person}{Nal Kalchbrenner}, \bibinfo{person}{Phil Blunsom},
  {and} \bibinfo{person}{Nando de Freitas}.} \bibinfo{year}{2014}\natexlab{}.
\newblock \showarticletitle{Modelling, visualising and summarising documents
  with a single convolutional neural network}.
\newblock \bibinfo{journal}{\emph{arXiv preprint arXiv:1406.3830}}
  (\bibinfo{year}{2014}).
\newblock


\bibitem[\protect\citeauthoryear{Devlin, Chang, Lee, and Toutanova}{Devlin
  et~al\mbox{.}}{2018}]%
        {devlin2018bert}
\bibfield{author}{\bibinfo{person}{Jacob Devlin}, \bibinfo{person}{Ming-Wei
  Chang}, \bibinfo{person}{Kenton Lee}, {and} \bibinfo{person}{Kristina
  Toutanova}.} \bibinfo{year}{2018}\natexlab{}.
\newblock \showarticletitle{Bert: Pre-training of deep bidirectional
  transformers for language understanding}.
\newblock \bibinfo{journal}{\emph{arXiv preprint arXiv:1810.04805}}
  (\bibinfo{year}{2018}).
\newblock


\bibitem[\protect\citeauthoryear{Dolan, Quirk, and Brockett}{Dolan
  et~al\mbox{.}}{2004}]%
        {dolan2004unsupervised}
\bibfield{author}{\bibinfo{person}{Bill Dolan}, \bibinfo{person}{Chris Quirk},
  {and} \bibinfo{person}{Chris Brockett}.} \bibinfo{year}{2004}\natexlab{}.
\newblock \showarticletitle{Unsupervised construction of large paraphrase
  corpora: Exploiting massively parallel news sources}. In
  \bibinfo{booktitle}{\emph{Proceedings of the 20th international conference on
  Computational Linguistics}}. Association for Computational Linguistics,
  \bibinfo{pages}{350}.
\newblock


\bibitem[\protect\citeauthoryear{Fan, Pang, Hou, Guo, Lan, and Cheng}{Fan
  et~al\mbox{.}}{2017}]%
        {fan2017matchzoo}
\bibfield{author}{\bibinfo{person}{Yixing Fan}, \bibinfo{person}{Liang Pang},
  \bibinfo{person}{JianPeng Hou}, \bibinfo{person}{Jiafeng Guo},
  \bibinfo{person}{Yanyan Lan}, {and} \bibinfo{person}{Xueqi Cheng}.}
  \bibinfo{year}{2017}\natexlab{}.
\newblock \showarticletitle{Matchzoo: A toolkit for deep text matching}.
\newblock \bibinfo{journal}{\emph{arXiv preprint arXiv:1707.07270}}
  (\bibinfo{year}{2017}).
\newblock


\bibitem[\protect\citeauthoryear{Fang, Tao, and Zhai}{Fang
  et~al\mbox{.}}{2004}]%
        {fang2004formal}
\bibfield{author}{\bibinfo{person}{Hui Fang}, \bibinfo{person}{Tao Tao}, {and}
  \bibinfo{person}{ChengXiang Zhai}.} \bibinfo{year}{2004}\natexlab{}.
\newblock \showarticletitle{A formal study of information retrieval
  heuristics}. In \bibinfo{booktitle}{\emph{Proceedings of the 27th annual
  international ACM SIGIR conference on Research and development in information
  retrieval}}. ACM, \bibinfo{pages}{49--56}.
\newblock


\bibitem[\protect\citeauthoryear{Fang and Zhai}{Fang and Zhai}{2005}]%
        {fang2005exploration}
\bibfield{author}{\bibinfo{person}{Hui Fang} {and} \bibinfo{person}{ChengXiang
  Zhai}.} \bibinfo{year}{2005}\natexlab{}.
\newblock \showarticletitle{An exploration of axiomatic approaches to
  information retrieval}. In \bibinfo{booktitle}{\emph{Proceedings of the 28th
  annual international ACM SIGIR conference on Research and development in
  information retrieval}}. \bibinfo{pages}{480--487}.
\newblock


\bibitem[\protect\citeauthoryear{Floyd and Fall}{Floyd and Fall}{1999}]%
        {Floyd1999Promoting}
\bibfield{author}{\bibinfo{person}{S. Floyd} {and} \bibinfo{person}{K. Fall}.}
  \bibinfo{year}{1999}\natexlab{}.
\newblock \showarticletitle{Promoting the use of end-to-end congestion control
  in the Internet}.
\newblock \bibinfo{journal}{\emph{IEEE/ACM Transactions on Networking}}
  \bibinfo{volume}{7}, \bibinfo{number}{4} (\bibinfo{year}{1999}),
  \bibinfo{pages}{458--472}.
\newblock


\bibitem[\protect\citeauthoryear{Guo, Fan, Ai, and Croft}{Guo
  et~al\mbox{.}}{2016}]%
        {Guo2016A}
\bibfield{author}{\bibinfo{person}{Jiafeng Guo}, \bibinfo{person}{Yixing Fan},
  \bibinfo{person}{Qingyao Ai}, {and} \bibinfo{person}{W.~Bruce Croft}.}
  \bibinfo{year}{2016}\natexlab{}.
\newblock \showarticletitle{A Deep Relevance Matching Model for Ad-hoc
  Retrieval}. In \bibinfo{booktitle}{\emph{Acm International on Conference on
  Information \& Knowledge Management}}.
\newblock


\bibitem[\protect\citeauthoryear{Hu, Lu, Li, and Chen}{Hu
  et~al\mbox{.}}{2014}]%
        {hu2014convolutional}
\bibfield{author}{\bibinfo{person}{Baotian Hu}, \bibinfo{person}{Zhengdong Lu},
  \bibinfo{person}{Hang Li}, {and} \bibinfo{person}{Qingcai Chen}.}
  \bibinfo{year}{2014}\natexlab{}.
\newblock \showarticletitle{Convolutional neural network architectures for
  matching natural language sentences}. In \bibinfo{booktitle}{\emph{Advances
  in neural information processing systems}}. \bibinfo{pages}{2042--2050}.
\newblock


\bibitem[\protect\citeauthoryear{Huang, He, Gao, Deng, Acero, and Heck}{Huang
  et~al\mbox{.}}{2013}]%
        {huang2013learning}
\bibfield{author}{\bibinfo{person}{Po-Sen Huang}, \bibinfo{person}{Xiaodong
  He}, \bibinfo{person}{Jianfeng Gao}, \bibinfo{person}{Li Deng},
  \bibinfo{person}{Alex Acero}, {and} \bibinfo{person}{Larry Heck}.}
  \bibinfo{year}{2013}\natexlab{}.
\newblock \showarticletitle{Learning deep structured semantic models for web
  search using clickthrough data}. In \bibinfo{booktitle}{\emph{Proceedings of
  the 22nd ACM international conference on Conference on information \&
  knowledge management}}. ACM, \bibinfo{pages}{2333--2338}.
\newblock


\bibitem[\protect\citeauthoryear{Kalchbrenner, Grefenstette, and
  Blunsom}{Kalchbrenner et~al\mbox{.}}{2014}]%
        {kalchbrenner2014convolutional}
\bibfield{author}{\bibinfo{person}{Nal Kalchbrenner}, \bibinfo{person}{Edward
  Grefenstette}, {and} \bibinfo{person}{Phil Blunsom}.}
  \bibinfo{year}{2014}\natexlab{}.
\newblock \showarticletitle{A convolutional neural network for modelling
  sentences}.
\newblock \bibinfo{journal}{\emph{arXiv preprint arXiv:1404.2188}}
  (\bibinfo{year}{2014}).
\newblock


\bibitem[\protect\citeauthoryear{Kelly et~al\mbox{.}}{Kelly
  et~al\mbox{.}}{2009}]%
        {kelly2009foundations}
\bibfield{author}{\bibinfo{person}{Diane Kelly} {et~al\mbox{.}}}
  \bibinfo{year}{2009}\natexlab{}.
\newblock \showarticletitle{Foundations and Trends{\textregistered} in
  Information Retrieval}.
\newblock \bibinfo{journal}{\emph{Foundations and Trends{\textregistered} in
  Information Retrieval}} \bibinfo{volume}{3}, \bibinfo{number}{1-2}
  (\bibinfo{year}{2009}), \bibinfo{pages}{1--224}.
\newblock


\bibitem[\protect\citeauthoryear{Kingma and Ba}{Kingma and Ba}{2014}]%
        {kingma2014adam}
\bibfield{author}{\bibinfo{person}{Diederik~P Kingma} {and}
  \bibinfo{person}{Jimmy Ba}.} \bibinfo{year}{2014}\natexlab{}.
\newblock \showarticletitle{Adam: A method for stochastic optimization}.
\newblock \bibinfo{journal}{\emph{arXiv preprint arXiv:1412.6980}}
  (\bibinfo{year}{2014}).
\newblock


\bibitem[\protect\citeauthoryear{Koh and Liang}{Koh and Liang}{2017}]%
        {koh2017understanding}
\bibfield{author}{\bibinfo{person}{Pang~Wei Koh} {and} \bibinfo{person}{Percy
  Liang}.} \bibinfo{year}{2017}\natexlab{}.
\newblock \showarticletitle{Understanding black-box predictions via influence
  functions}. In \bibinfo{booktitle}{\emph{Proceedings of the 34th
  International Conference on Machine Learning-Volume 70}}. JMLR. org,
  \bibinfo{pages}{1885--1894}.
\newblock


\bibitem[\protect\citeauthoryear{Li, Luong, and Jurafsky}{Li
  et~al\mbox{.}}{2015}]%
        {li2015hierarchical}
\bibfield{author}{\bibinfo{person}{Jiwei Li}, \bibinfo{person}{Minh-Thang
  Luong}, {and} \bibinfo{person}{Dan Jurafsky}.}
  \bibinfo{year}{2015}\natexlab{}.
\newblock \showarticletitle{A hierarchical neural autoencoder for paragraphs
  and documents}.
\newblock \bibinfo{journal}{\emph{arXiv preprint arXiv:1506.01057}}
  (\bibinfo{year}{2015}).
\newblock


\bibitem[\protect\citeauthoryear{Lin, Chen, and Yan}{Lin et~al\mbox{.}}{2013}]%
        {lin2013network}
\bibfield{author}{\bibinfo{person}{Min Lin}, \bibinfo{person}{Qiang Chen},
  {and} \bibinfo{person}{Shuicheng Yan}.} \bibinfo{year}{2013}\natexlab{}.
\newblock \showarticletitle{Network in network}.
\newblock \bibinfo{journal}{\emph{arXiv preprint arXiv:1312.4400}}
  (\bibinfo{year}{2013}).
\newblock


\bibitem[\protect\citeauthoryear{Mitra, Diaz, and Craswell}{Mitra
  et~al\mbox{.}}{2017}]%
        {mitra2017learning}
\bibfield{author}{\bibinfo{person}{Bhaskar Mitra}, \bibinfo{person}{Fernando
  Diaz}, {and} \bibinfo{person}{Nick Craswell}.}
  \bibinfo{year}{2017}\natexlab{}.
\newblock \showarticletitle{Learning to match using local and distributed
  representations of text for web search}. In
  \bibinfo{booktitle}{\emph{Proceedings of the 26th International Conference on
  World Wide Web}}. International World Wide Web Conferences Steering
  Committee, \bibinfo{pages}{1291--1299}.
\newblock


\bibitem[\protect\citeauthoryear{Nguyen, Rosenberg, Song, Gao, Tiwary,
  Majumder, and Deng}{Nguyen et~al\mbox{.}}{2016}]%
        {nguyen2016ms}
\bibfield{author}{\bibinfo{person}{Tri Nguyen}, \bibinfo{person}{Mir
  Rosenberg}, \bibinfo{person}{Xia Song}, \bibinfo{person}{Jianfeng Gao},
  \bibinfo{person}{Saurabh Tiwary}, \bibinfo{person}{Rangan Majumder}, {and}
  \bibinfo{person}{Li Deng}.} \bibinfo{year}{2016}\natexlab{}.
\newblock \showarticletitle{MS MARCO: a human-generated machine reading
  comprehension dataset}.
\newblock  (\bibinfo{year}{2016}).
\newblock


\bibitem[\protect\citeauthoryear{Olah, Mordvintsev, and Schubert}{Olah
  et~al\mbox{.}}{2017}]%
        {olah2017feature}
\bibfield{author}{\bibinfo{person}{Chris Olah}, \bibinfo{person}{Alexander
  Mordvintsev}, {and} \bibinfo{person}{Ludwig Schubert}.}
  \bibinfo{year}{2017}\natexlab{}.
\newblock \showarticletitle{Feature visualization}.
\newblock \bibinfo{journal}{\emph{Distill}} \bibinfo{volume}{2},
  \bibinfo{number}{11} (\bibinfo{year}{2017}), \bibinfo{pages}{e7}.
\newblock


\bibitem[\protect\citeauthoryear{Palangi, Deng, Shen, Gao, He, Chen, Song, and
  Ward}{Palangi et~al\mbox{.}}{2016}]%
        {palangi2016deep}
\bibfield{author}{\bibinfo{person}{Hamid Palangi}, \bibinfo{person}{Li Deng},
  \bibinfo{person}{Yelong Shen}, \bibinfo{person}{Jianfeng Gao},
  \bibinfo{person}{Xiaodong He}, \bibinfo{person}{Jianshu Chen},
  \bibinfo{person}{Xinying Song}, {and} \bibinfo{person}{Rabab Ward}.}
  \bibinfo{year}{2016}\natexlab{}.
\newblock \showarticletitle{Deep sentence embedding using long short-term
  memory networks: Analysis and application to information retrieval}.
\newblock \bibinfo{journal}{\emph{IEEE/ACM Transactions on Audio, Speech and
  Language Processing (TASLP)}} \bibinfo{volume}{24}, \bibinfo{number}{4}
  (\bibinfo{year}{2016}), \bibinfo{pages}{694--707}.
\newblock


\bibitem[\protect\citeauthoryear{Pang, Lan, Guo, Xu, Wan, and Cheng}{Pang
  et~al\mbox{.}}{2016}]%
        {pang2016text}
\bibfield{author}{\bibinfo{person}{Liang Pang}, \bibinfo{person}{Yanyan Lan},
  \bibinfo{person}{Jiafeng Guo}, \bibinfo{person}{Jun Xu},
  \bibinfo{person}{Shengxian Wan}, {and} \bibinfo{person}{Xueqi Cheng}.}
  \bibinfo{year}{2016}\natexlab{}.
\newblock \showarticletitle{Text Matching as Image Recognition.}. In
  \bibinfo{booktitle}{\emph{AAAI}}. \bibinfo{pages}{2793--2799}.
\newblock


\bibitem[\protect\citeauthoryear{Pennington, Socher, and Manning}{Pennington
  et~al\mbox{.}}{2014}]%
        {pennington2014glove}
\bibfield{author}{\bibinfo{person}{Jeffrey Pennington},
  \bibinfo{person}{Richard Socher}, {and} \bibinfo{person}{Christopher
  Manning}.} \bibinfo{year}{2014}\natexlab{}.
\newblock \showarticletitle{Glove: Global vectors for word representation}. In
  \bibinfo{booktitle}{\emph{Proceedings of the 2014 conference on empirical
  methods in natural language processing (EMNLP)}}.
  \bibinfo{pages}{1532--1543}.
\newblock


\bibitem[\protect\citeauthoryear{Qin and Liu}{Qin and Liu}{2013}]%
        {qin2013introducing}
\bibfield{author}{\bibinfo{person}{Tao Qin} {and} \bibinfo{person}{Tie-Yan
  Liu}.} \bibinfo{year}{2013}\natexlab{}.
\newblock \showarticletitle{Introducing LETOR 4.0 datasets}.
\newblock \bibinfo{journal}{\emph{arXiv preprint arXiv:1306.2597}}
  (\bibinfo{year}{2013}).
\newblock


\bibitem[\protect\citeauthoryear{Rennings, Moraes, and Hauff}{Rennings
  et~al\mbox{.}}{2019}]%
        {rennings2019axiomatic}
\bibfield{author}{\bibinfo{person}{Dani{\"e}l Rennings},
  \bibinfo{person}{Felipe Moraes}, {and} \bibinfo{person}{Claudia Hauff}.}
  \bibinfo{year}{2019}\natexlab{}.
\newblock \showarticletitle{An axiomatic approach to diagnosing neural ir
  models}. In \bibinfo{booktitle}{\emph{European Conference on Information
  Retrieval}}. Springer, \bibinfo{pages}{489--503}.
\newblock


\bibitem[\protect\citeauthoryear{Revaud, Weinzaepfel, Harchaoui, and
  Schmid}{Revaud et~al\mbox{.}}{2016}]%
        {revaud2016deepmatching}
\bibfield{author}{\bibinfo{person}{Jerome Revaud}, \bibinfo{person}{Philippe
  Weinzaepfel}, \bibinfo{person}{Zaid Harchaoui}, {and}
  \bibinfo{person}{Cordelia Schmid}.} \bibinfo{year}{2016}\natexlab{}.
\newblock \showarticletitle{Deepmatching: Hierarchical deformable dense
  matching}.
\newblock \bibinfo{journal}{\emph{International Journal of Computer Vision}}
  \bibinfo{volume}{120}, \bibinfo{number}{3} (\bibinfo{year}{2016}),
  \bibinfo{pages}{300--323}.
\newblock


\bibitem[\protect\citeauthoryear{Salton and Buckley}{Salton and
  Buckley}{1987}]%
        {Salton1987Term}
\bibfield{author}{\bibinfo{person}{Gerard Salton} {and}
  \bibinfo{person}{Christopher Buckley}.} \bibinfo{year}{1987}\natexlab{}.
\newblock \showarticletitle{Term-weighting approaches in automatic text
  retrieval}.
\newblock \bibinfo{journal}{\emph{Information Processing \& Management}}
  \bibinfo{volume}{24}, \bibinfo{number}{5} (\bibinfo{year}{1987}),
  \bibinfo{pages}{513--523}.
\newblock


\bibitem[\protect\citeauthoryear{Shen, He, Gao, Deng, and Mesnil}{Shen
  et~al\mbox{.}}{2014}]%
        {shen2014latent}
\bibfield{author}{\bibinfo{person}{Yelong Shen}, \bibinfo{person}{Xiaodong He},
  \bibinfo{person}{Jianfeng Gao}, \bibinfo{person}{Li Deng}, {and}
  \bibinfo{person}{Gr{\'e}goire Mesnil}.} \bibinfo{year}{2014}\natexlab{}.
\newblock \showarticletitle{A latent semantic model with convolutional-pooling
  structure for information retrieval}. In
  \bibinfo{booktitle}{\emph{Proceedings of the 23rd ACM International
  Conference on Conference on Information and Knowledge Management}}. ACM,
  \bibinfo{pages}{101--110}.
\newblock


\bibitem[\protect\citeauthoryear{Shrikumar, Greenside, and Kundaje}{Shrikumar
  et~al\mbox{.}}{2017}]%
        {shrikumar2017learning}
\bibfield{author}{\bibinfo{person}{Avanti Shrikumar}, \bibinfo{person}{Peyton
  Greenside}, {and} \bibinfo{person}{Anshul Kundaje}.}
  \bibinfo{year}{2017}\natexlab{}.
\newblock \showarticletitle{Learning important features through propagating
  activation differences}. In \bibinfo{booktitle}{\emph{Proceedings of the 34th
  International Conference on Machine Learning-Volume 70}}. JMLR. org,
  \bibinfo{pages}{3145--3153}.
\newblock


\bibitem[\protect\citeauthoryear{Shrikumar, Greenside, Shcherbina, and
  Kundaje}{Shrikumar et~al\mbox{.}}{2016}]%
        {shrikumar2016not}
\bibfield{author}{\bibinfo{person}{Avanti Shrikumar}, \bibinfo{person}{Peyton
  Greenside}, \bibinfo{person}{Anna Shcherbina}, {and} \bibinfo{person}{Anshul
  Kundaje}.} \bibinfo{year}{2016}\natexlab{}.
\newblock \showarticletitle{Not just a black box: Learning important features
  through propagating activation differences}.
\newblock \bibinfo{journal}{\emph{arXiv preprint arXiv:1605.01713}}
  (\bibinfo{year}{2016}).
\newblock


\bibitem[\protect\citeauthoryear{Simonyan, Vedaldi, and Zisserman}{Simonyan
  et~al\mbox{.}}{2013}]%
        {Simonyan2013Deep}
\bibfield{author}{\bibinfo{person}{Karen Simonyan}, \bibinfo{person}{Andrea
  Vedaldi}, {and} \bibinfo{person}{Andrew Zisserman}.}
  \bibinfo{year}{2013}\natexlab{}.
\newblock \showarticletitle{Deep inside convolutional networks: Visualising
  image classification models and saliency maps}.
\newblock \bibinfo{journal}{\emph{arXiv preprint arXiv:1312.6034}}
  (\bibinfo{year}{2013}).
\newblock


\bibitem[\protect\citeauthoryear{Springenberg, Dosovitskiy, Brox, and
  Riedmiller}{Springenberg et~al\mbox{.}}{2014}]%
        {springenberg2014striving}
\bibfield{author}{\bibinfo{person}{Jost~Tobias Springenberg},
  \bibinfo{person}{Alexey Dosovitskiy}, \bibinfo{person}{Thomas Brox}, {and}
  \bibinfo{person}{Martin Riedmiller}.} \bibinfo{year}{2014}\natexlab{}.
\newblock \showarticletitle{Striving for simplicity: The all convolutional
  net}.
\newblock \bibinfo{journal}{\emph{arXiv preprint arXiv:1412.6806}}
  (\bibinfo{year}{2014}).
\newblock


\bibitem[\protect\citeauthoryear{Sundararajan, Taly, and Yan}{Sundararajan
  et~al\mbox{.}}{2017}]%
        {sundararajan2017axiomatic}
\bibfield{author}{\bibinfo{person}{Mukund Sundararajan}, \bibinfo{person}{Ankur
  Taly}, {and} \bibinfo{person}{Qiqi Yan}.} \bibinfo{year}{2017}\natexlab{}.
\newblock \showarticletitle{Axiomatic attribution for deep networks}. In
  \bibinfo{booktitle}{\emph{Proceedings of the 34th International Conference on
  Machine Learning-Volume 70}}. JMLR. org, \bibinfo{pages}{3319--3328}.
\newblock


\bibitem[\protect\citeauthoryear{Touretzky, Mozer, and Hasselmo}{Touretzky
  et~al\mbox{.}}{1996}]%
        {touretzky1996advances}
\bibfield{author}{\bibinfo{person}{David~S Touretzky},
  \bibinfo{person}{Michael~C Mozer}, {and} \bibinfo{person}{Michael~E
  Hasselmo}.} \bibinfo{year}{1996}\natexlab{}.
\newblock \bibinfo{booktitle}{\emph{Advances in Neural Information Processing
  Systems 8: Proceedings of the 1995 Conference}}. Vol.~\bibinfo{volume}{8}.
\newblock \bibinfo{publisher}{Mit Press}.
\newblock


\bibitem[\protect\citeauthoryear{Wan, Lan, Guo, Xu, Pang, and Cheng}{Wan
  et~al\mbox{.}}{2016a}]%
        {wan2016deep}
\bibfield{author}{\bibinfo{person}{Shengxian Wan}, \bibinfo{person}{Yanyan
  Lan}, \bibinfo{person}{Jiafeng Guo}, \bibinfo{person}{Jun Xu},
  \bibinfo{person}{Liang Pang}, {and} \bibinfo{person}{Xueqi Cheng}.}
  \bibinfo{year}{2016}\natexlab{a}.
\newblock \showarticletitle{A Deep Architecture for Semantic Matching with
  Multiple Positional Sentence Representations.}. In
  \bibinfo{booktitle}{\emph{AAAI}}, Vol.~\bibinfo{volume}{16}.
  \bibinfo{pages}{2835--2841}.
\newblock


\bibitem[\protect\citeauthoryear{Wan, Lan, Xu, Guo, Pang, and Cheng}{Wan
  et~al\mbox{.}}{2016b}]%
        {wan2016match}
\bibfield{author}{\bibinfo{person}{Shengxian Wan}, \bibinfo{person}{Yanyan
  Lan}, \bibinfo{person}{Jun Xu}, \bibinfo{person}{Jiafeng Guo},
  \bibinfo{person}{Liang Pang}, {and} \bibinfo{person}{Xueqi Cheng}.}
  \bibinfo{year}{2016}\natexlab{b}.
\newblock \showarticletitle{Match-srnn: Modeling the recursive matching
  structure with spatial rnn}.
\newblock \bibinfo{journal}{\emph{arXiv preprint arXiv:1604.04378}}
  (\bibinfo{year}{2016}).
\newblock


\bibitem[\protect\citeauthoryear{Wang, Hamza, and Florian}{Wang
  et~al\mbox{.}}{2017}]%
        {wang2017bilateral}
\bibfield{author}{\bibinfo{person}{Zhiguo Wang}, \bibinfo{person}{Wael Hamza},
  {and} \bibinfo{person}{Radu Florian}.} \bibinfo{year}{2017}\natexlab{}.
\newblock \showarticletitle{Bilateral multi-perspective matching for natural
  language sentences}.
\newblock \bibinfo{journal}{\emph{arXiv preprint arXiv:1702.03814}}
  (\bibinfo{year}{2017}).
\newblock


\bibitem[\protect\citeauthoryear{Xiong, Dai, Callan, Liu, and Power}{Xiong
  et~al\mbox{.}}{2017}]%
        {xiong2017end}
\bibfield{author}{\bibinfo{person}{Chenyan Xiong}, \bibinfo{person}{Zhuyun
  Dai}, \bibinfo{person}{Jamie Callan}, \bibinfo{person}{Zhiyuan Liu}, {and}
  \bibinfo{person}{Russell Power}.} \bibinfo{year}{2017}\natexlab{}.
\newblock \showarticletitle{End-to-end neural ad-hoc ranking with kernel
  pooling}. In \bibinfo{booktitle}{\emph{Proceedings of the 40th International
  ACM SIGIR conference on research and development in information retrieval}}.
  ACM, \bibinfo{pages}{55--64}.
\newblock


\bibitem[\protect\citeauthoryear{Zobel and Moffat}{Zobel and Moffat}{1998}]%
        {Zobel1998Exploring}
\bibfield{author}{\bibinfo{person}{Justin Zobel} {and}
  \bibinfo{person}{Alistair Moffat}.} \bibinfo{year}{1998}\natexlab{}.
\newblock \showarticletitle{Exploring the similarity space}.
\newblock \bibinfo{journal}{\emph{Acm Sigir Forum}} \bibinfo{volume}{32},
  \bibinfo{number}{1} (\bibinfo{year}{1998}), \bibinfo{pages}{18--34}.
\newblock


\end{thebibliography}

\end{document}